\documentclass[a4paper,fleqn]{cas-dc}

\usepackage{physics}
\usepackage[numbers]{natbib}
\usepackage{float}
\usepackage[utf8]{inputenc}
\hypersetup{
    colorlinks=true,
    linkcolor=blue,
    citecolor=blue,
    urlcolor=blue,
}

\def\tsc#1{\csdef{#1}{\textsc{\lowercase{#1}}\xspace}}
\tsc{WGM}
\tsc{QE}

\begin{document}
\let\WriteBookmarks\relax
\def\floatpagepagefraction{1}
\def\textpagefraction{.001}

\shorttitle{Emergence of Transport Regimes from the Axial Interfacial Gradients in Nanopores}   

\shortauthors{PS,DP}  

\title{Emergence of Transport Regimes from the Axial Field-Induced Interfacial Gradients in Uniform Surface Potential Nanopores}



%

\author[1]{Pramodt Srinivasula}[orcid=0009-0002-6002-2394]

\cormark[1]


\ead{pramodt.research@gmail.com}


\credit{Conceptualization, Methodology, Analysis, Manuscript preparation and Review}

\affiliation[1]{organization={ElectroSoft Labs LLP},
            city={Mumbai},
            postcode={400063}, 
            state={Maharashtra},
            country={India}}


 \author[2]{Doyel Pandey}[orcid=0000-0002-8331-5424]




\credit{Methodology, Simulations, Manuscript preparation}

\affiliation[2]{organization={School of Mathematical \& Statistical Sciences, IIT Mandi},
            postcode={175075}, 
            state={Himachal Pradesh},
            country={India}}



\begin{abstract}
Gate-modulated nanopores have emerged as a promising platform for achieving ion selectivity and ionic current rectification (ICR) with the advantage of active field-based control. However, the mechanistic origin of these experimentally reported phenomena, arising from electrostatic coupling between the prescribed radial pore surface potential and the axial transmembrane electric field, remains insufficiently understood.
Here, using coupled Poisson--Nernst--Planck and Navier--Stokes simulations supported by asymptotic analysis, we show that a uniform surface potential inherently interacts with the axial driving field to generate a three-dimensional, axially nonuniform electric double layer (EDL). This field-induced EDL heterogeneity effectively mimics a linear axial variation in zeta potential, breaking translational symmetry within an otherwise uniform pore.
As a result, the system exhibits coupled electrokinetic responses, including ion selectivity, ionic current rectification, and non-canonical electroosmotic flow, all governed by a single asymmetry parameter $\alpha$ derived from the EDL structure. Critical transitions occur at specific values of $\alpha$; in particular, at $\alpha\!=\!0$, the EDL becomes axially antisymmetric, leading to reversal of ion selectivity, significant ICR and the emergence of a peculiar negative electroosmotic flow rectification accompanied by internal vortical structures.
These findings establish the electrostatic mechanism for axial symmetry breaking as the underlying principle for transport in voltage-gated nanopores, enabling a unified framework for designing tunable electrokinetic functionalities beyond geometry- and chemistry-based strategies.
\end{abstract}





\maketitle

\section{Introduction}\label{sec:Intro}

Voltage-gated nanopores have become popular owing to advances in solid-state nanofabrication originally developed for nanoelectronics. Key enablers include two-dimensional membranes (e.g., graphene, Hexagonal boron nitride, MoS$_2$), conductive materials (e.g., Au, CNTs, ITO), and high-$\kappa$ dielectrics (e.g., Al$_2$O$_3$, TiO$_2$), integrated using techniques such as electron-beam lithography and atomic layer deposition \cite{nam2010sub,liu2016slowing,robin2023long}. Surround-gate or ionic field-effect transistor geometries created by conformally layering dielectrics and metals around nanopores enable precise electrostatic modulation of pore walls via capacitive coupling \cite{nam2010sub}. This voltage gating approach provides dynamic control over ion transport and electroosmotic flow, offering greater flexibility than other methods such as, temperature or salinity gradient modulation \cite{yeh2012slowing,he2013mechanism}. As a result, gated nanopores have found broad applications across neuromorphic nanofluidic logic \cite{robin2023long,guan2011field}, biosensing and DNA analysis \cite{liu2016slowing,sugimoto2015dna}, and energy conversion technologies \cite{tsutsui2024gate,lei2025ultrahigh}.

Solid-state nanochannels equipped with gate electrodes or redox-active coatings can function as ionic diodes and transistors by modulating surface charge and electroosmotic flow (EOF). Gate bias dynamically tunes ionic current, selectivity, and transport polarity. For example, \citet{guan2011field} demonstrated a field-effect nanofluidic diode with an asymmetrically positioned gate along the length of the pore, enabling directional switching of ionic conduction. \citet{fuest2015three} realized a three-state ionic switch where varying gate voltage yielded forward, off, or reverse current. Gate-controlled chemical coatings such as PEDOT allowed tunable ion selectivity between cations, anions, or neutral states \citep{perez2017all}, while redox-active coatings operated as ionic field-effect transistors \citep{laucirica2019redox}. 

Voltage gating is emerging as a powerful tool to enhance membrane selectivity and energy harvesting. In sub--2 nm graphene lamellar membranes, an applied $\pm 0.5$ V gate accelerates ion diffusion 4 to 7 fold \cite{cheng2018low}. A similar approach with conducting membranes significantly improved NaCl rejection to 98.6\% while doubling water flux, overcoming the selectivity--permeance tradeoff \cite{zhang2025simultaneous}. Gated nanopores have also been used in salinity-gradient power generation, reaching power densities up to 15 W$\cdot$m$^{-2}$, ahead of the economical feasibility limit \cite{tsutsui2024gate,xiao2019ion}. These results underscore how field-effect modulation enables dynamic control of ionic transport for desalination and energy conversion \cite{ren2018voltage,laucirica2019redox,guan2011field}.

Recent investigations illustrated the underlying ion selectivity and current rectification behavior that arises in the gated nanopores. \citet{tsutsui2024gate} realized uniform pore surface potentials by embedding platinum gate electrodes in the nanopores of SiO$_x$ membrane and leveraged the resulting non-monotonous current-voltage variation as a rectification for harvesting osmotic power. \citet{ak2024electrostatic} used stacked graphene nanopores were used as voltage-gated nanopores using copper foil tape around the pore, and showed a clear trend of the gate-voltage modulated current rectification behavior and further compared relative selectivity of cations from multiple electrolytes of different valencies. \citet{chen2025electrostatically} used trilayer graphene to prescribe a surface potential to the nanopore and indicated the current variation with the potential deviates from the classical linear trend of uniformly charged surfaces. 

Ionic current rectification (ICR) and eletroosmotic flow (EOF) modulation are a key phenomena for many of such nanopore applications. ICR occurs when asymmetries, either geometric or material-based or electrostatic, cause differential ion accumulation under reversed biases. Conical pores with charged walls exhibit diode-like behavior due to asymmetric ion enrichment and depletion \cite{siwy2002fabrication,laohakunakorn2015electroosmotic,trivedi2022ion}. Various nanopore shapes (trumpet, cigar, hourglass, dumbbell) were also proposed to generate rich EOF profiles and tunable rectification under voltage \citep{chuang2023influence, khosravikia2023quantitative}, although they remain complex to be realized compared to voltage-gated nanopores. Similarly, patterned surface charges can induce rectification without geometric asymmetry achieved using channels with regions of nonuniform or oppositely charged surface segments  \cite{karnik2007rectification,kiy2023highly,zhang2024modulation}. Janus nanopores with bipolar charge distributions also exhibit asymmetric conduction \cite{montes2022ionic}. Even EOF is perturbed with a prescribed nonuniform zeta potential along the length of the pore, highlighting the critical role of charge nonuniformity \cite{fu2003analysis}.

Symmetry breaking and diode-like behavior can arise even in geometrically symmetric systems using asymmetric operating parameters or electrolytes. For example, a ion concentration or salinity gradient across a uniform silica nanochannel induces ionic current rectification by promoting ion accumulation in one bias direction and depletion in the other \cite{deng2014effect}. Since electroosmotic flow (EOF) depends on ionic conductivity, such gradients also cause flow rectification, with stronger EOF in the forward direction \cite{experton2017ion}. Also, valancy-asymmetric electrolytes can lead to a net EOF \cite{pandey2024net}. In systems with symmetric geometry, operating and material parameters, internal asymmetry is needed to attain rectification behaviors, which was achieved using voltage-gating, resulting in the useful nanotechnological applications mentioned above.

In modeling, voltage-gated nanopores are treated with electrostatic boundaries where the wall or zeta potential is externally controlled via capacitive coupling to a gate, rather than fixed by surface chemistry. \citet{liu2010descreening} used a simple approximation of a gate-controlled surface potential of the pore which was demonstrated to be sufficient for the phenomenological insights. This approach, grounded in planar electrostatic models, allows gate voltage to modulate the electric double layer (EDL) and electroosmotic flow (EOF) via Poisson--Nernst--Planck and Stokes equations \cite{hu2012field,schoch2008transport}. While 1D models and molecular simulations have captured EDL and EOF scaling \cite{lian2017non,wang2014evaluation}, such analyses break down under higher-dimensional confinement where electrohydrodynamic interactions become inherently nonuniform \cite{bhattacharyya2005electro,hu2012field}. Notably, ion and fluid transport phenomena reported in nanopore experiments and applications are predominantly characterized under steady-state conditions, supporting the adequacy of steady-state modeling for their analysis.

An axial field interaction with a uniform gate potential was reported to effectively result in descreening effect in the nanopore \cite{liu2010descreening}. However, its contribution to breaking the symmetry, and hence in modulating the ICR and EOF directly relevant for the applications was not fully explored. In this work, we investigate this configuration using coupled Poisson--Nernst--Planck and Navier--Stokes simulations. By systematically varying the transmembrane potential relative to the gate voltage, we elucidate the mechanisms governing the onset of axial asymmetry in the electric field and EDL structure, thereby characterizing the resulting emergent transport regimes.

\section{Mathematical modeling}\label{subsec:Model}

\begin{figure}[]
    \centering
    \includegraphics[width=0.95\linewidth]{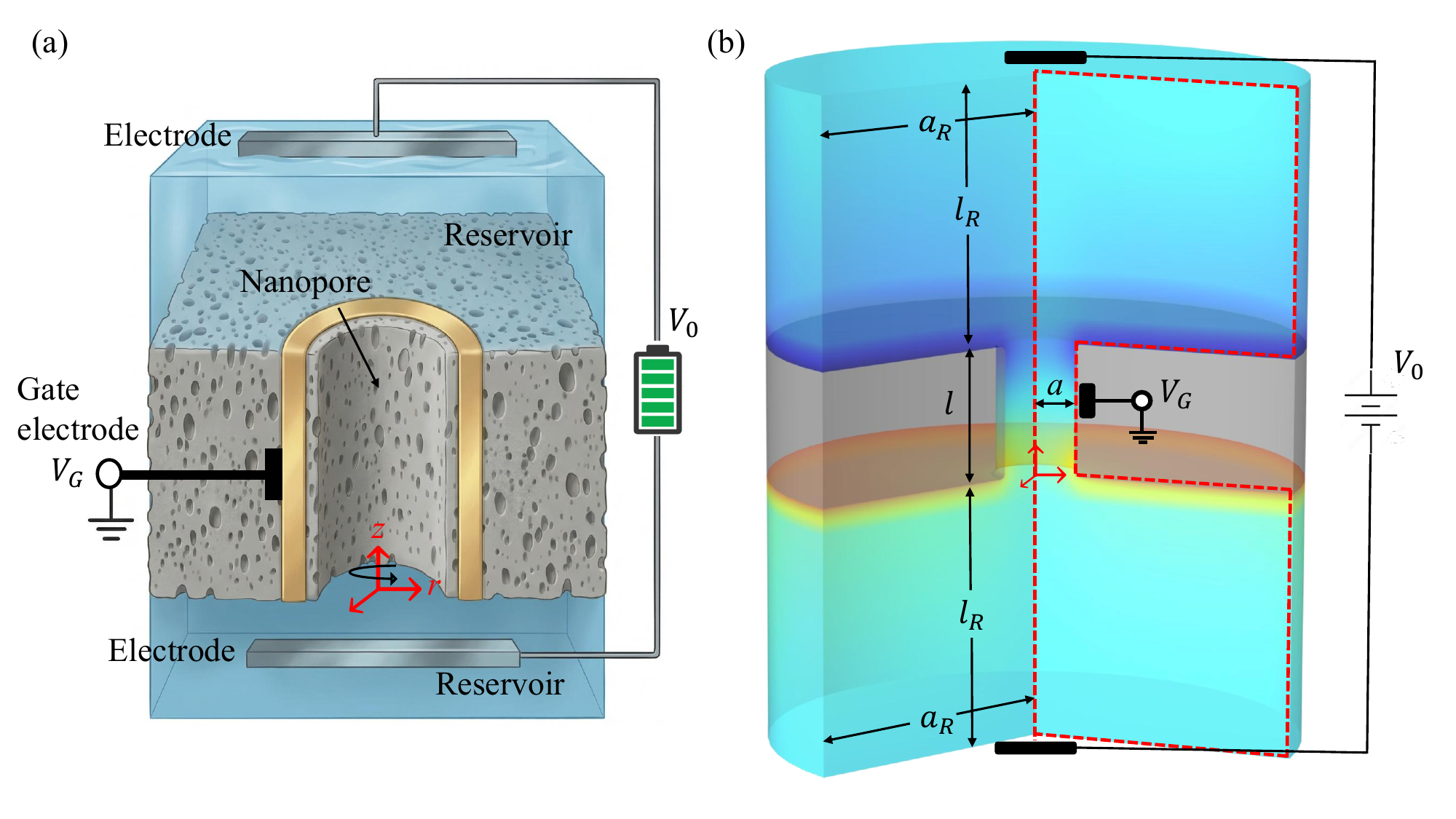}
    \caption{(a) Physical and (b) mathematical model of a cylindrical nanopore of radius $a$, length $l$, connecting two identical reservoirs of dimension $(a_R \times l_R)$, filled with incompressible, binary, monovalent electrolyte solution. An asymmetric cylindrical coordinate system ($r,z$) is considered with the origin placed at the junction of the nanopore and the bottom reservoir. The computational domain is highlighted by red dashed lines. A gate electrode is engraved of the nanopore wall which possess an gate voltage $V_G=-25$ mV. On the other hand, an external voltage $V_0$ is applied across the nanopore in the axial direction through the electrodes placed in both the reservoirs. EOF and ion transfer takes place through the nanopore form one reservoir to another.}
    \label{fig:1_Illustration}
\end{figure}
We formulate a steady-state continuum model to describe ionic transport and electroosmotic flow (EOF) through a cylindrical nanopore of radius $a (=\!50$ nm) and length $l (=\!1~\mu$m), connecting two identical electrolyte reservoirs. The nanopore wall is electrically coupled to a gated electrode (Cf. Fig.~\ref{fig:1_Illustration}), which capacitively imposes a near uniform surface potential corresponding to a gate voltage $V_G (=\!-25$ mV). In addition, a transmembrane voltage $V_0$ is applied across the two reservoirs via external electrodes, driving ionic transport through the pore. The system is filled with a binary, symmetric, monovalent electrolyte ($i=\{+,-\}$ with valences $z_+\!=\!+1$ and $z_-\!=\!-1$) of bulk concentration $c_0$, modeled as a Newtonian and incompressible fluid. 
Each reservoir is represented by a domain of size $a_R \times l_R$, as illustrated in Fig.~\ref{fig:1_Illustration}(b). Owing to the axisymmetric geometry of the system, the governing equations are solved in a two-dimensional cylindrical coordinate framework $(r,z)$, with the origin located at the junction between the nanopore and the lower reservoir.
The electrolyte properties are assumed constant, with density $\rho$ and viscosity $\mu$ corresponding to those of water at $T (=300$ K). The medium is characterized by a uniform dielectric permittivity $\varepsilon = \varepsilon_0 \varepsilon_r$, where $\varepsilon_0$ is the vacuum permittivity and $\varepsilon_r$ is the relative permittivity of the electrolyte.

The nanopore walls may possess an intrinsic surface charge density $\sigma$, while a transmembrane potential difference $V_0$ is externally applied between the reservoirs to drive ionic transport. In classical electrokinetic models, the wall electrostatics are typically prescribed via zeta potential $\zeta$, a  fixed surface potential parameter independent of the electrolyte properties, imposed as a Dirichlet boundary condition \cite{masliyah2006electrokinetic,sherwood1995electrophoresis,park2006eddies,sadeghi2019electroosmotic,pandey2025contribution}. Under such assumptions, the interfacial electrokinetics remain decoupled from the externally applied axial field, allowing the electrostatic problem to be decomposed into a Poisson equation governing the equilibrium electric double layer and a Laplace equation describing the bulk potential distribution associated with the transmembrane bias.

In contrast, when a gate potential is imposed in conjunction with the inherent surface charge, the surface electrostatics become intrinsically coupled to the applied axial field. In this case, the electrostatic potential cannot be decomposed into independent components; instead, the gate-induced surface potential and the transmembrane field interact nonlinearly, jointly determining the spatially varying structure of the electric double layer within the nanopore. The governing equations for this axisymmetric setup are formulated below. 

\paragraph{Electrostatics:} The electrostatic potential $\phi$ satisfies a single Poisson equation
\begin{equation}\label{eq:Poisson_phi}
\nabla\cdot\!\left(\varepsilon_0\varepsilon_r\nabla\phi\right)=-\rho_e(r,z),
\end{equation}
where the space-charge density $\rho_e$ dependents on the spatial coordinates $(r,z)$ is related to the ionic species concentrations $c_i(r,z)$ via the Faraday constant $F$ through
\begin{equation}
\rho_e(r,z) = F \sum_{i=+,-} z_i c_i(r,z).
\end{equation}

An electric field $E_g\!\approx \!-(V_G-\phi_s)/t_G$ is created across a thin dielectric layer of the membrane of thickness $t_G$ and relative permittivity $\varepsilon_{rm}$ sandwiched between the pore inner wall with potential $\phi_s$ and the implanted metal gating at potential $V_G$ (illustrated in Fig. \ref{fig:1_Illustration}(a)). This adds a capacitive charge, $\sigma_G\!=\!\varepsilon_0 \varepsilon_{rm} E_g$ onto the pore inner wall in addition to the inherent surface charge ($\sigma$) arising from the surface chemistry of the membrane pores. Hence, a mixed boundary condition arises at the pore inner walls as, 
\begin{align}
  \!\!\!\!\!\!\!\!\!\!\!\!\!  -\varepsilon_0\varepsilon_r \nabla\phi\cdot\mathbf{n}=\sigma -\varepsilon_0 \varepsilon_{rm} \frac{V_G-\phi}{t_G},
\quad \forall (r=a), z\!\in(0,l)
\label{eq:MixedBC}
\end{align}
Also, simplified cases of a \textit{fixed surface charge }(FSC) condition for a uniform pore surface charge or \textit{fixed surface potential} (FSP) condition for a uniform electric potential can be formulated as below \citep{srinivasula2025dipolar}, to be used when applicable, as discussed later.
\begin{equation}\label{eq:FSC_BC}
-\varepsilon_0\varepsilon_r\,\nabla\phi\cdot\mathbf{n}=\sigma,
\quad \forall\, (r=a),\; z\in(0,l),
\end{equation}
or
\begin{equation}\label{eq:FSP_BC}
\phi=V_G,
\quad \quad \forall\, (r=a),\; z\in(0,l)
\end{equation}
Here, $\mathbf{n}$ denotes the outward unit normal to the corresponding surfaces. 

The nanopore is subjected to the transmembrane potential drop $V_0$ across the nanopore between the upper and lower reservoirs as 
\begin{equation}\label{eq:reservoir_phi_BC1}
\phi=V_0,
\quad \forall\, r\in(0,a_R),\quad z=l+l_R
\end{equation}
and
\begin{equation}\label{eq:reservoir_phi_BC2}
\phi=0, \quad \forall\, r\in(0,a_R),\quad z=-l_R.
\end{equation}
Since, the space charge within the large reservoirs is negligible, a nominal axial electric field of magnitude of the order $E_0\!=\!V_0/l$ is generated across the pore length.   
On all the other surfaces, no electrical charge condition is applied as $\mathbf{n}\cdot\nabla\phi = 0$. 

\paragraph{Ion transport:} The ionic species concentrations are governed by the steady-state Nernst-Planck equations, which account for diffusion, electromigration in the electric potential $\phi(r,z)$, and advection by the fluid velocity field $\mathbf{u}$,
\begin{equation}
\begin{split}
\nabla \cdot \mathbf{N}_i &= 0, \\
\mathbf{N}_i &= c_i \mathbf{u} - D_i \nabla c_i 
- D_i \frac{z_i F}{RT}\, c_i \nabla \phi ,
\label{eq:NPeq}
\end{split}
\end{equation}
where $\mathbf{N}_i$ and $D_i$ denote the total flux and diffusion coefficient of $i^{th}$ species, respectively and $R$ is the universal gas constant.
The outer computational boundaries of the large upper and lower reservoirs are maintained at a constant electrolyte concentration,
\begin{equation}\label{eq:c_0Reservoirs}
c_i = c_0, \quad \forall\, r\in(0,a_R),\quad z = -l_R \ \text{and} \ z = l+l_R.
\end{equation}
No-flux boundary condition $\mathbf{n}\!\cdot\! \mathbf{N}_i \!=\! 0$ is imposed on all impermeable surfaces of the membrane including the pore inner walls ($\forall r\!=\!a, z \!\in\!(0,l)$) and membrane-reservoir boundary $(\forall r\in (a,a_R), z = 0 \ \text{and} \ z = l)$. Also, a zero net flux of ions arises on the radial computational boundaries of the reservoirs $(r=a_R, \forall z\in (-l_R,0) \cup (l,l+l_R)$ owing to the symmetric distribution of nanopores across the membrane. 

\paragraph{Fluid dynamics:} 
The fluid motion of the electrolyte is governed by the Stokes equation coupled with the continuity equation for the velocity field $\mathbf{u}(r,z)$ and pressure $p(r,z)$. The flow is driven by an electric body force arising from the local space-charge density $\rho_e$, 
\begin{align}
-\nabla p + \mu \nabla \cdot \left( \nabla \mathbf{u} + \nabla \mathbf{u}^T \right) - \rho_e \nabla \phi &= \mathbf{0}, \label{eq:Stokes}\\
\nabla \cdot \mathbf{u} &= 0. \label{eq:continuity}
\end{align}
No-slip boundary conditions are imposed on all membrane surfaces.
\begin{align}
\!\!\!\!\!\!\!\!\!\!\!\!\mathbf{u}\! =\! \mathbf{0}, \, \forall\, (r \!=\! a,\! z \!\in\! (0,l)) \! \cup \! \left(r \!\in\! (a,a_R),\!  z \!= \!0 \ \text{and} \ z \!=\! l \right)
\end{align}
To represent the large reservoirs, a uniform reference pressure is prescribed at the upper and lower boundaries of the computational domain with no net normal stress.
\begin{align}
n \cdot ( p\mathbf{I} + \mu \left( \nabla \mathbf{u} + \nabla \mathbf{u}^T \right) =0, \quad p = 0, \quad \quad  \quad\nonumber\\
 \!\!\!\!\!\!\!\!\!\!\!\!\forall\, r \in (0,a_R), \quad z = -l_R \ \text{and} \ z = l + l_R
\end{align}
To model the nanopore as a representative of a membrane comprising a large array of uniformly distributed nanopores, symmetry-like boundary conditions are imposed on the lateral reservoir boundaries enforcing vanishing tangential viscous stress and zero normal velocity.
\begin{align}
&\mu \left( \nabla \mathbf{u} + (\nabla \mathbf{u})^{T} \right) \cdot (\mathbf{I} - \mathbf{n}\mathbf{n}) = 0,\nonumber\\
&\mathbf{u} \cdot \mathbf{n} = 0 \quad \forall\, r = a_R, \; z \in (-l_R,0) \cup (l,l+l_R).
\end{align}

By exploiting axisymmetry, the three-dimensional problem is reduced to a two-dimensional formulation over a symmetric half-plane shown in Fig. \ref{App:Numerical_details}, with the below symmetry conditions.
\begin{align}
  \!\!\!\!\!\!\!\!\!\!\!\!\!\!\!  \pdv{\phi}{r}\! =\!0,  \pdv{c_i}{r}\!=\!0,  \pdv{u_z}{r}\!=\!u_r\!=0, \,\,\forall  (r\!=\!0),\! z\! \in\! (-l_R,l\!+\!l_R)
\end{align}
A schematic representation of all boundary conditions is provided in Fig.~\ref{fig:App_BC} in Appendix~\ref{App:Numerical_details} for clarity.

\paragraph{Numerical implementation:}\label{subsec:NumericalImplementation}

The coupled Poisson--Nernst--Planck equations \eqref{eq:Poisson_phi}, \eqref{eq:NPeq}, together with the Stokes equation \eqref{eq:Stokes} and the continuity equation \eqref{eq:continuity}, are solved numerically under the boundary conditions described above. The computations are performed using the finite-element method implemented in COMSOL Multiphysics. 
A non-uniform, structured mesh is employed to adequately resolve the length scales of mechanics present in the system, particularly within the thin electrical double layer near membrane surfaces and in the larger bulk electrolyte. The mesh is refined in regions of steep gradients while maintaining computational efficiency in the bulk domain. A comprehensive grid-independence study is conducted, yielding a grid convergence index (GCI) below $ 1\%$, thereby ensuring that the numerical solutions are insensitive to further mesh refinement. Details of the numerical formulation, solver settings, mesh distribution, and grid-independence analysis are provided in Appendix~\ref{App:Numerical_details}.

Upon confirming the grid convergence test to ensure the numerical precision, the results are verified for the mass and species conservation via flux and current continuities along the length of the pore. No spurious steady state fluxes are observed with zero transmembrane potential.

\subsection{Validation of the numerical model}\label{App:Validation}

The surface electrostatic condition for a voltage-gated nanopore (Eq.~\eqref{eq:MixedBC}) can be expressed in terms of the dimensionless parameter $\beta = \kappa t_G \varepsilon_r / \varepsilon_{rm}$ as
\begin{align}
\phi_s = V_G - \beta \left( \pdv{\phi_s}{r} + \sigma \frac{\varepsilon_r}{\varepsilon_{0}} \right).
\end{align}
Here $\kappa = \sqrt{2e_0^2 c_0/(\varepsilon_0 \varepsilon_r k_\mathrm{B} T)}$ is the inverse Debye length with the elementary charge $e_0$ and Boltzmann constant $k_B$. The three terms on the right-hand side correspond to contributions from a fixed surface potential (FSP), a Stern-layer response \citep{bazant2004diffuse,srinivasula2025dipolar}, and a fixed surface charge (FSC), respectively. 

For systems with a thin dielectric separation between the gating electrode and the pore wall ($t_G \ll 1$), $\beta$ is typically small. This is relevant for highly polarizable nanopores, such as wetted MXene membranes \citep{wang2014evaluation}, where $\beta \sim 0.05$--$0.5$ for bulk electrolyte concentrations in the range of $1$--$100$ mM. In addition, for systems with negligible intrinsic surface charge (e.g., SiO$_2$-coated nanopores in \citet{tsutsui2024gate}), the FSC contribution is minimal. Even in cases where $\beta \!=\! \mathcal{O}(1)$, the combined Stern-layer and FSC effects in cylindrical nanopores with symmetric electrolytes are not expected to produce pronounced non-primitive transport behaviors such as selectivity or rectification within the linear potential regime. 

\begin{figure}[]
    \centering
    \includegraphics[width=0.47\linewidth]{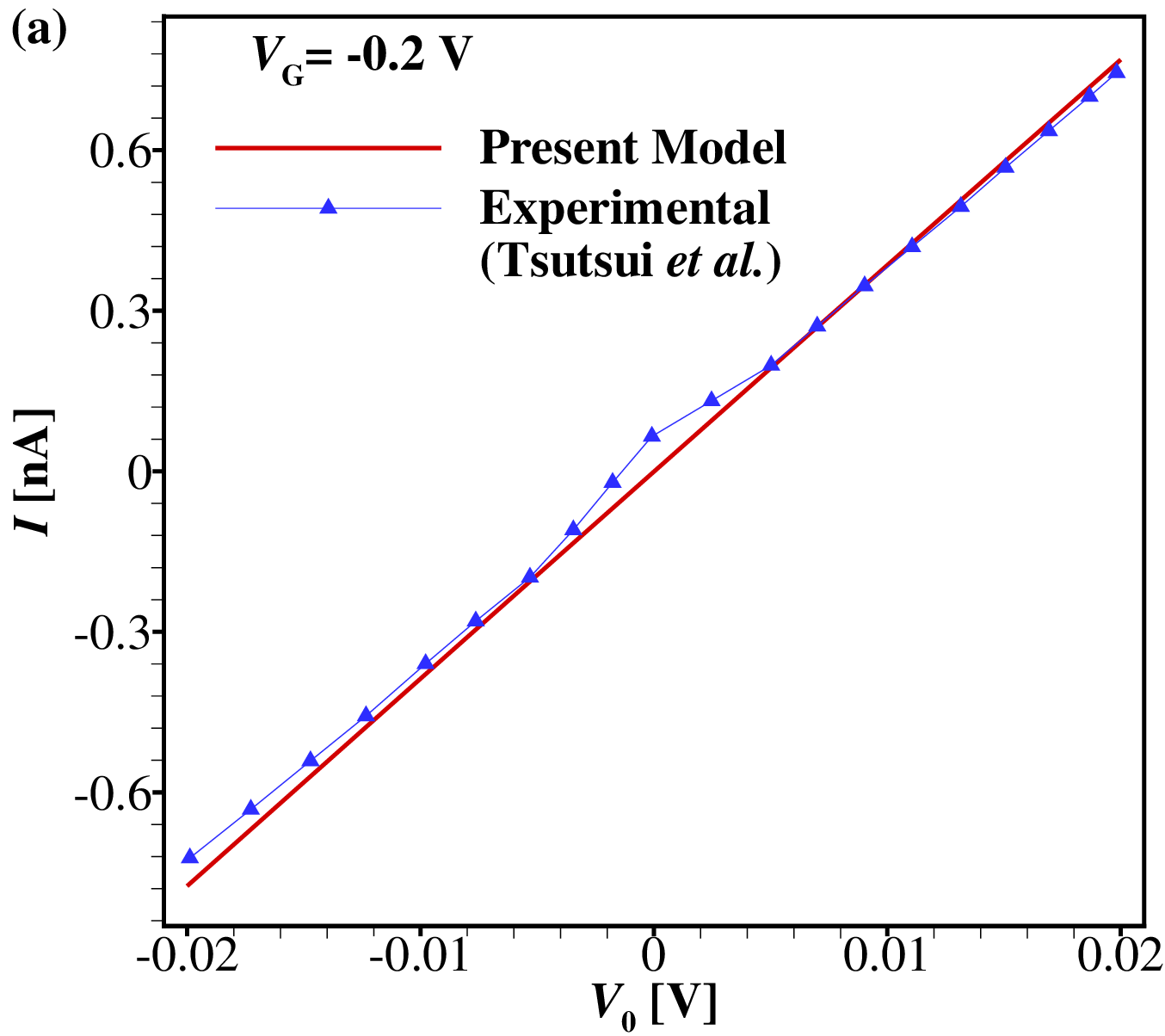}
    \includegraphics[width=0.47\linewidth]{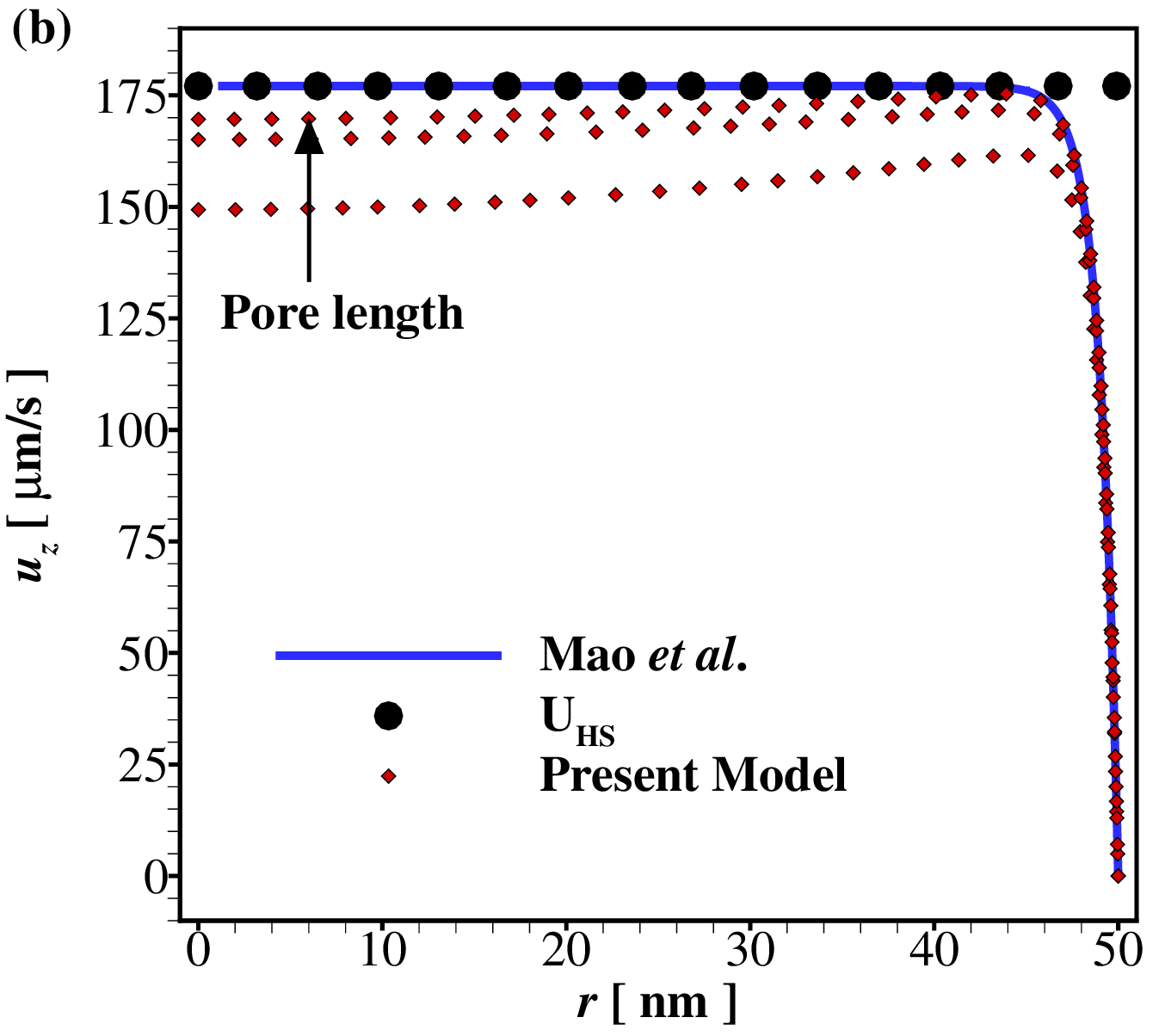}
    \caption{\textit{Validation of numerical method.} (a) Comparison of $I-V$ results of the present model using FSP condition (Eq. \ref{eq:FSP_BC}) with the experimental work of Tsutsui \textit{et al}. \cite{tsutsui2024gate}. (b) Axial velocity of the fluid at a cross-section of the nanopore of radius $50$ nm away from the pore-reservoir junction. Blue solid line denotes the analytical electroosmotic velocity of an infinitely long cylindrical channel \cite{mao2014electro}, and circular symbols denote the Helmoltz-Smoluchowski slip velocity. Diamond shaped symbols represents the axial velocity of the fluid for nanopore lengths $l\!=\!1\mu$m, 2.5$\mu$m and 5$\mu$m.}
    \label{fig:App_validation}
\end{figure}

Motivated by these considerations, the present study focuses on isolating the role of the FSP contribution in governing the emergence of non-primitive transport phenomena.
The numerical predictions of the FSP model are compared against experimental measurements of ionic current in voltage-gated nanopores reported by \citet{tsutsui2024gate}. As shown in Fig.~\ref{fig:App_validation}(a), the current--voltage characteristics are compared for a nanopore of radius $a = 150$ nm and length $l = 90$ nm, with a gate voltage $V_G = -200$ mV, over $V_0 \in [-20,\,20]$ mV from the experiments. The electrolyte is a NaCl solution of bulk concentration $13.7$ mM, corresponding to $\kappa a \!\approx\! 57$, and the intrinsic surface charge is negligible.

The model predictions show quantitative agreement with the experimental data over the considered voltage range. In the experimental configuration, both the pore walls and the membrane boundary adjacent to one reservoir are gated. However, the present result (Fig. \ref{fig:App_validation}(a)) indicate that applying the FSP condition solely on the pore wall is sufficient to capture the observed current response at low to moderate potentials, suggesting that the essential electrostatic mechanisms are adequately represented in the model. In this study, lower gate voltages are employed to suppress strongly nonlinear EDL effects, enabling isolation of the fundamental mechanisms governing non-primitive transport phenomena.

On the other hand, a clear trend of EOF with voltage-gating is rarely reported, although it is studied with a fixed surface charge condition. Helmholtz-Smoluchowski estimation for a nominal value of EOF velocity magnitude, $u_{HS}\!=\!\varepsilon_0\varepsilon_r \zeta E_0/\mu$ is a baseline which estimates the velocity for infinitely thin EDL. Mao \textit{et al}. \cite{mao2014electro} extended this to estimate the velocity for thin EDL near charged surfaces of an infinitely long nanochannel. Fig.~\ref{fig:App_validation}(b) shows the variation of the axial EOF velocity with the radial coordinate in the nanopore for different nanopore lengths, while the surface charge $\sigma$ is prescribed such that the zeta potential is $-25$ mV. As the pore aspect ratio increases, the results converge towards the analytical estimates of Mao \textit{et al}. \cite{mao2014electro} based on the EOF velocity in an infinite nanochannel, i.e., without the end effects.

To isolate and elucidate the role of externally controlled gate potential, the present study considers this effect through a uniform fixed surface potential independently from intrinsic fixed surface charge. Accordingly, cases with FSP and FSC are evaluated separately and compared, allowing a clear distinction between their respective mechanisms influencing the transport behavior.

\section{Asymptotic analysis of a uniform surface potential nanopore}\label{sec:Asymptotic}
In the asymptotic analysis, we restrict attention to the interior of the cylindrical nanopore domain, $\{0 \leq r \leq a, \; 0 \leq z \leq l\}$ occupied by the monovalent electrolyte of equal ionic diffusivities ($D_+\!=\! D_-$), and bulk concentration $c_0$. A uniform FSP $V_G$ is imposed on the cylindrical wall at $r \!=\! a$, excluding the corners which are constrained by end potentials $0$ and $V_0$ are prescribed at the lower ($z\!=\!0$) and upper ($z\!=\!l$) boundaries, respectively.

\subsection{Equivalent zeta potential and electric field}
Under the Debye--H\"uckel approximation, $\left| e\phi / k_\mathrm{B}T \right| \ll 1$, the electrostatic potential $\phi(r,z)$ satisfies the linearized Poisson--Boltzmann equation,
\begin{equation}
\frac{1}{r}\frac{\partial}{\partial r}
\!\left(r\frac{\partial\phi}{\partial r}\right)
+
\frac{\partial^2\phi}{\partial z^2}
-
\kappa^2\phi
= 0.
\label{eq:DHcyl}
\end{equation}

The boundary conditions are specified as follows. On the cylindrical wall,
\begin{equation}
\phi(a,z) = V_G, \qquad z \in (\lambda, l - \lambda),
\label{eq:radialBC}
\end{equation}
while insulating conditions are imposed near the pore ends,
\begin{equation}
\nabla \phi(a,z) \cdot \mathbf{n} = 0, 
\qquad z \in (0,\lambda) \cup (l-\lambda,l),
\end{equation}
where $\lambda \ll l$ is a small buffer region introduced to regularize corner singularities. Axisymmetry requires $\partial \phi/\partial r = 0$ at $r = 0$.

\paragraph{Solution decomposition}
To facilitate analytical treatment, the potential is decomposed as
\begin{equation}
\phi(r,z) = \psi(z) + \varphi(r,z),
\label{eq:decomp}
\end{equation}
where $\psi(z)$ satisfies the one-dimensional Laplace equation,
\begin{equation}
\frac{d^2 \psi}{dz^2} = 0,
\end{equation}
subject to $\psi(0)=0$ and $\psi(l)=V_0$. This yields
\begin{equation}
\psi(z) = \frac{V_0}{l} z.
\end{equation}
The remaining component $\varphi(r,z)$ captures deviations induced by the wall boundary condition and satisfies
\begin{equation}
\frac{1}{r}\frac{\partial}{\partial r}
\!\left(r\frac{\partial\varphi}{\partial r}\right)
+
\frac{\partial^2\varphi}{\partial z^2}
-
\kappa^2 \varphi = 0,
\end{equation}
with axisymmetry and boundary conditions
\[
\frac{\partial \varphi}{\partial r}(0,z) = 0, 
\quad 
\varphi(a,z) = V_G - \psi(z), 
\quad 
\varphi(r,0) = \varphi(r,l) = 0.
\]

\paragraph{Series solution}
Seeking separable solutions of the form $\varphi(r,z) = R(r)Z(z)$ with the small buffer region ( small $\lambda$) leads to the eigenfunction expansion
\begin{align}
\varphi(r,z) &=
\sum_{n=1}^{\infty}
b_n \frac{I_0(\alpha_n r)}{I_0(\alpha_n a)}
\sin\!\left(\frac{n\pi z}{l}\right), \\
b_n &=
\frac{2}{l}
\int_{\lambda}^{l-\lambda}
\left( V_G - \frac{V_0\xi}{l} \right)
\sin\!\left(\frac{n\pi \xi}{l}\right)\, d\xi,
\label{eq:bnseries}
\end{align}
where $\alpha_n = \sqrt{\kappa^2 + \frac{n^2\pi^2}{l^2}}$, and $I_0$ denotes the modified Bessel function of the first kind. By combining these components, the complete electrostatic potential is therefore
\begin{equation}
\phi(r,z) = \frac{V_0}{l} z 
+ \sum_{n=1}^{\infty}
b_n \frac{I_0(\alpha_n r)}{I_0(\alpha_n a)}
\sin\!\left(\frac{n\pi z}{l}\right).
\label{eq:psi(r,z)}
\end{equation}

\paragraph{Equivalent local zeta potential}
The equivalent local zeta potential at any crosssection along the length of the pore is obtained by matching the radial potential drop to that of the one-dimensional (radial) Debye--H\"uckel solution in a cylindrical pore, $\phi(r,z)\!=\!\zeta_{eq} I_0(\kappa r)/I_0(\kappa a)$, yielding $\zeta_{\mathrm{eq}}(z)$.
\begin{equation}
\zeta_{\mathrm{eq}}(z) = 
\frac{\phi(a,z) - \phi(0,z)}{1 - 1/I_0(\kappa a)}
\label{eq:zeta_eq_def}
\end{equation}
This provides a reduced measure of the local electrostatic structure along the length of the nanopore.
Substituting Eq.~\eqref{eq:psi(r,z)} together with Eq.~\eqref{eq:bnseries} with the imposed surface potential $V_G$ within small buffer zone approximation, we obtain
\begin{equation}
\zeta_{\mathrm{eq}}(z)
=
\frac{
V_G - \dfrac{V_0 z}{l}
- \displaystyle\sum_{n=1}^{\infty}
\frac{b_n}{I_0(\alpha_n a)}
\sin\!\left(\frac{n\pi z}{l}\right)
}{
1 - \dfrac{1}{I_0(\kappa a)}
}.
\label{eq:zeta-dirichlet-explicit}
\end{equation}
Despite the spatially uniform fixed surface potential, $\zeta_{\mathrm{eq}}$ exhibits a pronounced axial variation, reflecting the coupled influence of the gate potential on the radial boundaries and the axial transmembrane electric field on the 3-dimensional EDL structure. However, the overall influence of this EDL structure is quantified by defining an axial average of the equivalent local zeta potential $\langle \zeta_{eq}\rangle$ across the length of the pore, evaluated in Appendix~\ref{sec:append_zeta_finiteEDL}.
\begin{equation}
\langle \zeta_{eq} \rangle = \frac{1}{l} \int_0^l \zeta_{\mathrm{eq}}(z)\,dz 
= V_G \left(1 - \frac{V_0}{2V_G}\right)
\label{eq:zeta_av}
\end{equation}

\paragraph{Axial electric field}
The axial electric field is obtained from the electrostatic potential as $\mathbf{E}_z=E_z\mathbf{\hat{z}}$, with magnitude $E_z = -\partial \phi/\partial z$, yielding from Eq.~\eqref{eq:psi(r,z)} as
\begin{equation}
E_z(r,z)
=
-\frac{V_0}{l}
-
\sum_{n=1}^{\infty}
\frac{n\pi}{l} \, b_n
\frac{I_0(\alpha_n r)}{I_0(\alpha_n a)}
\cos\!\left(\frac{n\pi z}{l}\right).
\label{eq:axial_Ez}
\end{equation}
This expression shows that the axial electric field is inherently non-uniform, exhibiting coupled axial and radial variations induced by the interaction between the transmembrane potential and the imposed surface potential.

\subsection{Thin EDL limit}\label{subsec:Thin}
\begin{figure}[]
    \centering
    \includegraphics[width=0.48\linewidth]{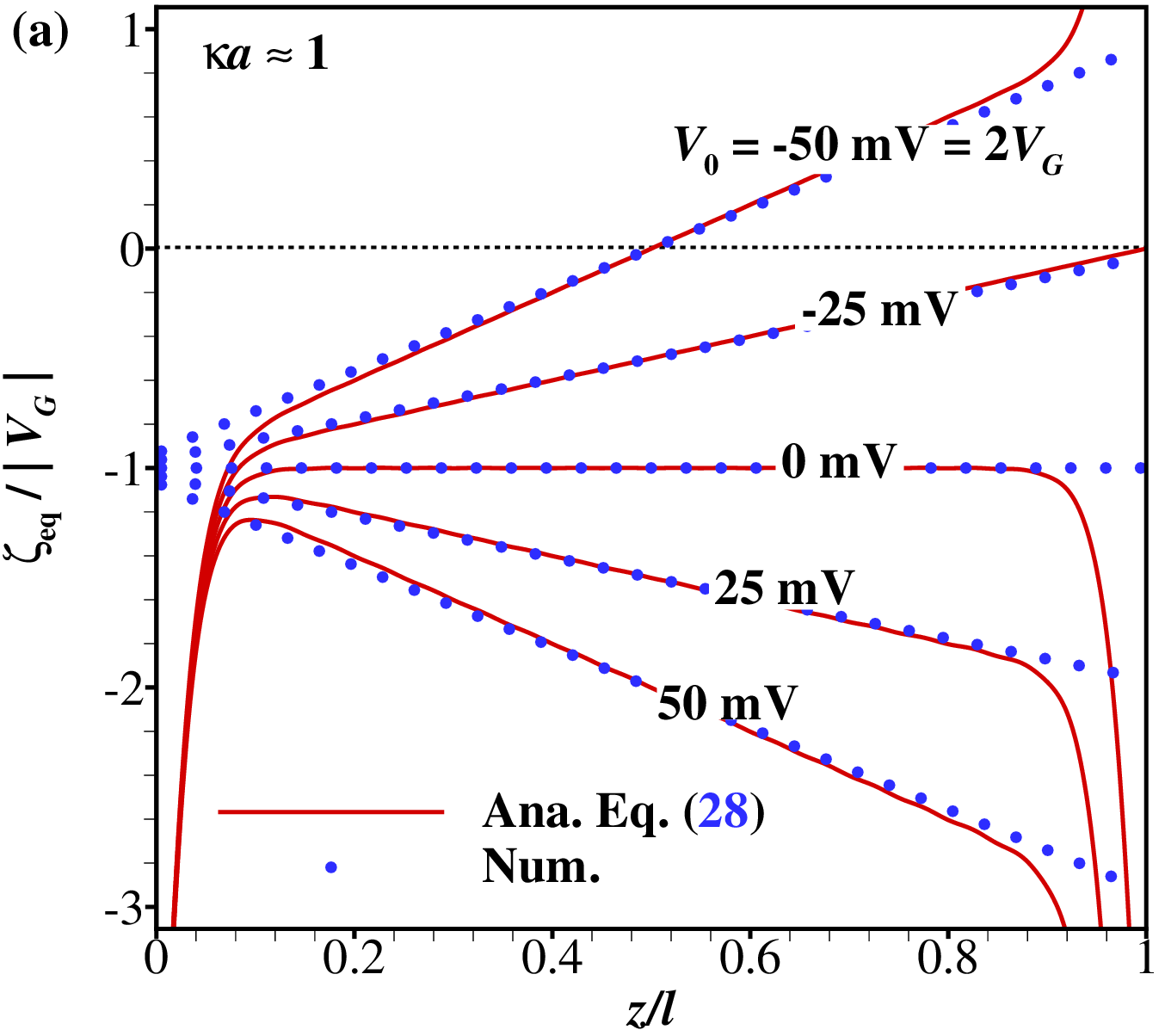}
    \includegraphics[width=0.48\linewidth]{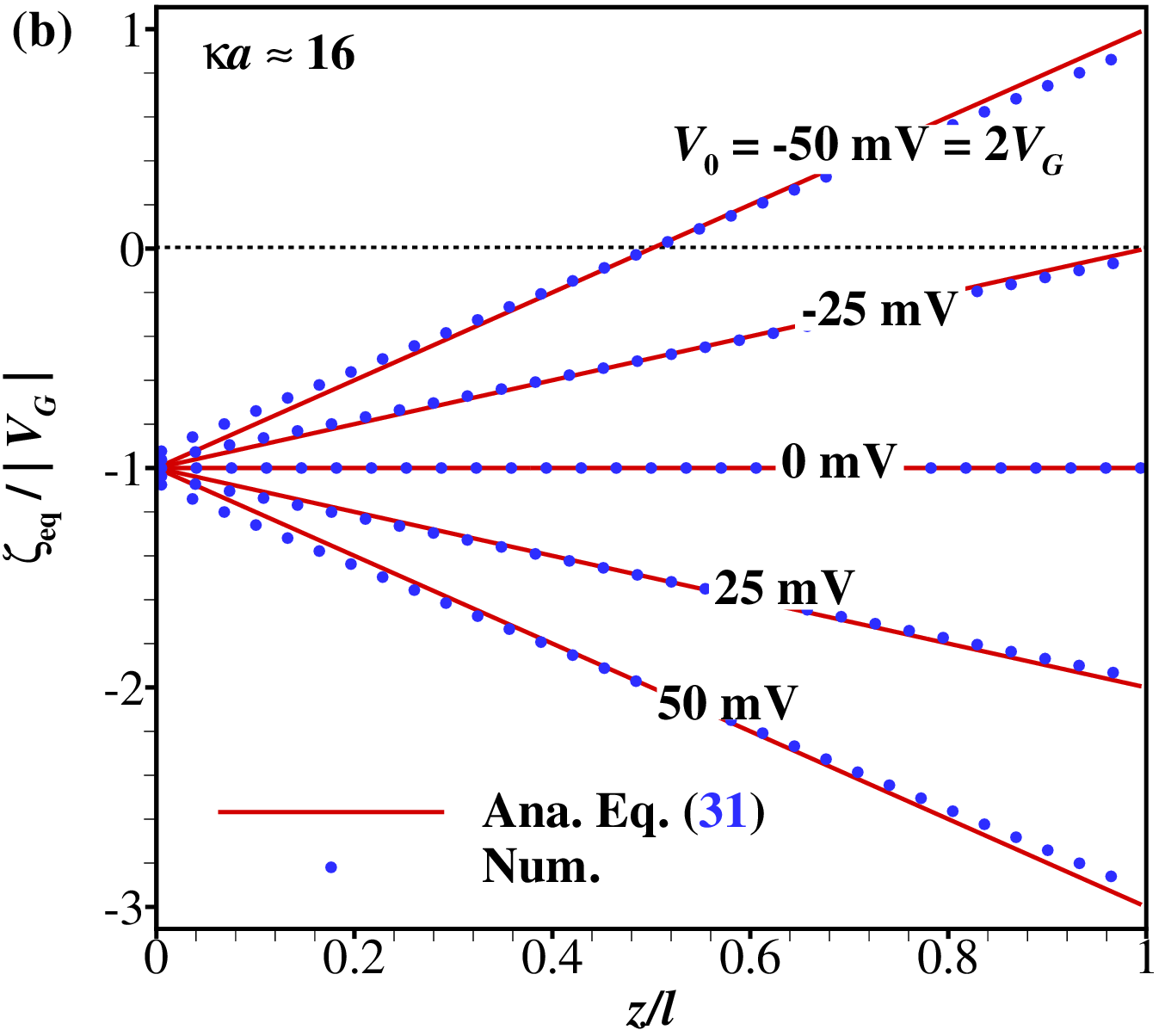}
    \caption{Non dimensional equivalent zeta potential along the axial direction of the nanochannel for  (a) $\kappa a=1.03$ Eq.~\eqref{eq:zeta-dirichlet-explicit}, (b) $\kappa a=16.2$ Eq.~\eqref{eq:zeta-dirichlet-explicit2} and different values of $V_0=\pm 25,\pm 50, 0$ mV. Other parameters are fixed at gate potential $V_G=-25$ mV, nanochannel length $l=1~\mu$m.}
    \label{fig:1D_zeta_gate_vs_V0}
\end{figure}

In the thin electrical double layer (EDL) limit, $\kappa a \gg 1$, the influence of the prescribed surface potential is confined to a narrow region adjacent to the pore wall and does not penetrate significantly into the pore interior. Consequently, the potential away from the pore walls is well approximated by the bulk linear profile arising from the nominal transmembrane field. For a long pore ($\alpha_n \!\approx \! \kappa$) and thin EDL $(I_0(\kappa a)\!\rightarrow\!\infty)$, the Eq.~\eqref{eq:psi(r,z)} gives $\phi(r,z)\! \simeq \!V_0 z/l$ outside for $r\! \gg \! \kappa^{-1}$. Under this limit, the equivalent local zeta potential from Eq.~\eqref{eq:zeta_eq_def} simplifies to
\begin{equation}
\zeta_{\mathrm{eq}}^{\mathrm{thin}}(z) = V_G - \frac{V_0 z}{l}.
\label{eq:zeta-dirichlet-explicit2}
\end{equation}
This explicitly indicates a linear variation along the pore length governed by the contribution from both the gate potential and the imposed transmembrane bias. 

In addition, the axial electric field in the thin EDL limit, from Eq.~\eqref{eq:axial_Ez} is $E_z^{thin}\!=\!-E_0 \!=\!-V_0/l $ as $\kappa a \!\rightarrow \! \infty$.

\subsection{Space charge density distribution}

At steady state, the EDL attains local equilibrium in the radial direction, with small radial advection, such that the radial ionic flux vanishes. Under this condition, the ionic concentrations follow a Boltzmann distribution, which can be expressed in terms of the equivalent zeta potential. Within the Debye--H\"uckel approximation, the resulting space charge density is given by
\begin{equation}
\rho_e (r,z) = - \varepsilon_0 \varepsilon_r \kappa^2 \, \zeta_{\mathrm{eq}}(z) \, \frac{ I_0(\kappa r)}{I_0(\kappa a)},
\label{eq:space_chrarge_analytical}
\end{equation}
demonstrating that the space charge distribution inherits an explicit axial dependence through $\zeta_{\mathrm{eq}}(z)$, thereby capturing the axial non-uniformity of the EDL structure within the nanopore.

\subsection{Electroosmotic velocity}
In the complete model with the reservoirds, the two outer computational boundaries of the reservoirs are maintained at constant pressures, while the fluid can move in or outwards through the open boundary. Since the reservoirs are very large compared to the pore, for the asymptotic analysis, we can approximate constant pressures at the two ends of the cylindrical pore. Owing to the mass conservation within the pore, the flow rate through the pore us uniform at any cross-section across the length of the pore, $Q(z)\!=\!\int_0^a 2\pi r dr\,u_z\!=\!Q_0$.
For a long pore, we can estimate the velocity and pressure ($p$) fields using a lubrication approximation $ \epsilon \!=\!a/l\! \ll \! 1 $. Then the $r-$momentum equation reduced to $\partial p/\partial r=0 $, while the axial momentum equation reduces to
\begin{equation}
\mu \nabla_r^2 u_z = \frac{dp}{dz} - \rho_e(r,z)\, E_z,
\end{equation}
where $\nabla_r^2 \!=\! \frac{1}{r} \frac{\partial}{\partial r}\left( r \frac{\partial}{\partial r} \right)$, $\rho_e(r,z)$ is given by Eq.~\eqref{eq:space_chrarge_analytical}. 

Solving this equation using superposition of pressure-driven and electroosmotic contributions, and enforcing regularity at the axis, yields
\begin{equation}
u_z(r,z)= \frac{1}{4\mu} \frac{dp}{dz} \! (r^2 \! - \! a^2) \! + \!\frac{\varepsilon_0 \varepsilon_r}{\mu}
\!\left[ \frac{I_0(\kappa r)}{I_0(\kappa a)} \! - \!1 \right]\! E_z \zeta_{\mathrm{eq}}(z). \label{eq:uz_in_prGrad}
\end{equation}
The flow rate through the pore hence consists of two components from the pressure-driven and EOF components as, $Q_0=Q_p+Q_e$, with 
\begin{align}
    Q_p &= -\frac{\pi a^4}{8 \mu} \pdv{p}{z}\\
    Q_e &= \frac{\varepsilon_0 \varepsilon_r}{\mu} E_z \zeta_{eq}(z) \pi a^2 \left( \frac{2 I_i(\kappa a)}{\kappa a I_0(\kappa a)} -1 \right)
\end{align}
Further, using the two reservoir conditions, $p(0)\!=\!p(l)\!=\!p_0$, we have $\int_0^l (dp/dz)\,dz \!=\!0$. This gives the pressure gradient in terms of the local deviation of equivalent zeta potentials from the average value across the pore, as  
\begin{equation}
\frac{dp}{dz} =
\frac{8 \varepsilon_0 \varepsilon_r E_z}{a^2}
\left[\frac{2}{\kappa a} \frac{I_1(\kappa a)}{I_0(\kappa a)} -1 \right]
\left( \zeta_{\mathrm{eq}}(z) - \langle \zeta_{eq} \rangle \right),
\label{eq:dpdz}
\end{equation}
with $\langle \zeta_{eq} \rangle$ obtained from Eq.~\eqref{eq:zeta_av}.

Substituting this expression back into the velocity field yields the complete axial velocity profile involving the axial field $E_z(r,z)$ that can vary spatially in both axial and radial directions. 
\begin{align}
\!\!\!\!\!\!\!\!\!\!\!\!\!\!u_z(r,z)
&=
\frac{\varepsilon_0 \varepsilon_r E_z(r,z)}{\mu}
\left[
\zeta_{\mathrm{eq}}(z)\left(\frac{I_0(\kappa r)}{I_0(\kappa a)} - 1\right)
\right. \nonumber \\
&\quad \left.
 -2\!
\left(\!\zeta_{\mathrm{eq}}(z)\!-\!\langle \zeta_{eq} \rangle \right)\!\!
\left(\!1 \!-\! \frac{r^2}{a^2}\!\right)\!\!\left(\!\frac{2}{\kappa a}\frac{I_1(\kappa a)}{I_0(\kappa a)} \!-\! 1\!\!\right)\!
\right]
\end{align}
Hence the cross-sectionally averaged mean velocity $u_m\!=\!(2/a^2) \int_0^a\!r u_z\!(r,z)\, dr$ is obtained that resembles the Helmholtz-Smoluchowski formula, however with the spatially varying local field and globally averaged equivalent zeta potentials.
\begin{align}
u_{m} & =
\frac{\varepsilon_0 \varepsilon_r}{\mu} E_z(r,z) \langle \zeta_{eq} \rangle 
\left[\frac{2}{\kappa a} \frac{I_1(\kappa a)}{I_0(\kappa a)}-1\right]
\label{eq:velocity_av}
\end{align}
In the thin EDL limit $\kappa a \gg1$, the radial coordinate dependence of the axial field is eliminated, and also the term $2I_1(\kappa a)/( \kappa a I_0(\kappa a))$ vanishes. Hence, using Eq.~\eqref{eq:zeta_av}, we obtain,
\begin{align}
u_{m}^{thin}  = \frac{\varepsilon_0 \varepsilon_r}{\mu l}
 V_0 V_G \left(1 - \frac{V_0}{2V_G}\right),
\label{eq:velocity_av_thin}
\end{align}
indicating a nonlinear dependence of the mean velocity on the applied potential.
\begin{figure*}[]
    \centering
    \includegraphics[width=0.9\linewidth]{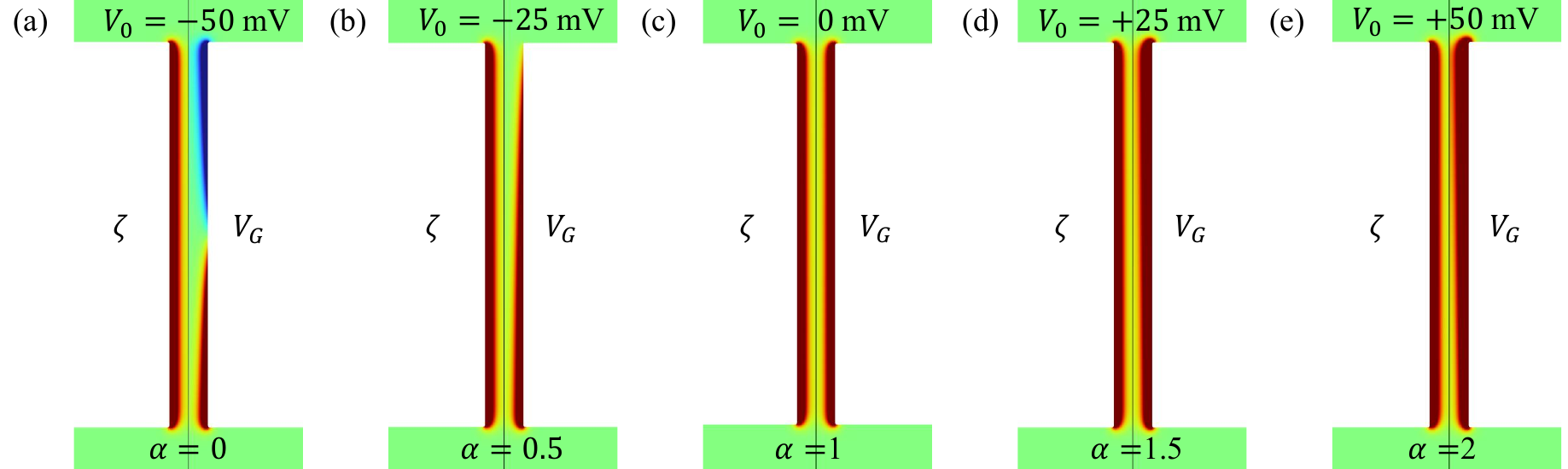}  
    \includegraphics[width=0.05\linewidth]{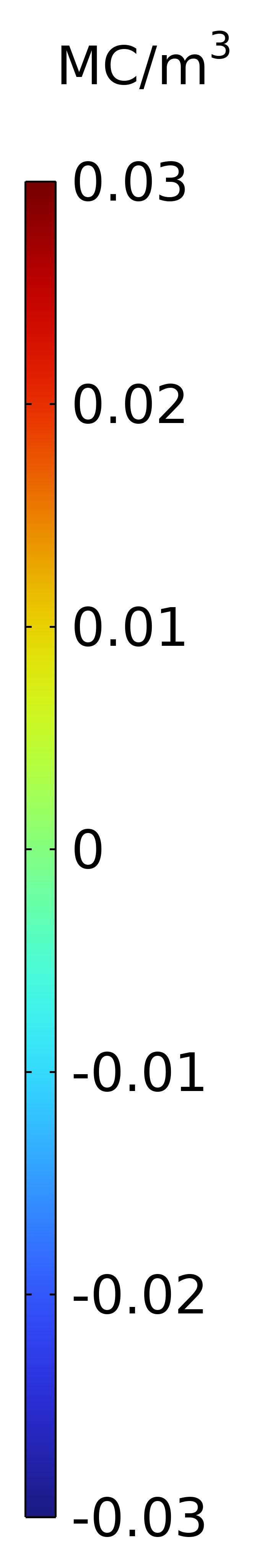}
    \caption{(a--e) Space charge density distributions within the nanopore at $c_0\!=\!1$ mM for different transmembrane potentials $V_0$ (corresponding $\alpha$ values indicated). Left: fixed surface charge (FSC) with $\zeta\!=\!-25$ mV; right: fixed surface potential (FSP) with $V_G\!=\!-25$ mV. Here pore length $l\!=\!1\mu$m and radius $a\!=\!50$ nm.}
    \label{fig:2D_zeta_eq}
\end{figure*}

\section{Axial EDL Nonuniformity}\label{subsec:AxialGradients}

\subsection{Equivalent zeta potential variation}

Despite a spatially uniform surface potential along the pore walls, a pronounced axial nonuniformity in the EDL structure emerges. This is quantified through the axial variation of the equivalent zeta potential, $\zeta_{\mathrm{eq}}(z)$, shown in Fig.~\ref{fig:1D_zeta_gate_vs_V0}, which serves as an effective one-dimensional representation of a fundamentally higher-dimensional restructuring of the EDL. Both numerical simulations and asymptotic analysis in small buffer zone limit ($\lambda \!\rightarrow\!0$) consistently reveal a linear variation of $\zeta_{\mathrm{eq}}$ along the pore length, observed for both moderately thick ($\kappa a = 1.03$) and thin ($\kappa a = 16.2$) EDLs (Fig.~\ref{fig:1D_zeta_gate_vs_V0}(a,b)).
The slope of this axial variation decreases systematically with increasing transmembrane potential $V_0$ over the range $[-2V_G,\,2V_G]$, for a fixed wall potential $V_G=-25$ mV. Notably, $\zeta_{\mathrm{eq}}$ locally exceeds or falls below the imposed surface potential.

This nonuniformity arises from electrostatic coupling between the axial electric field (imposed by $V_0$) and the radial field (modulated by $V_G$). Consider a case where the axial potential varies continuously between the two reservoir boundary values ($0,V_0$) encapsulating the fixed radial surface potential ($0<V_G<V_0$). Consequently, there exists an axial position where the potential on the axis and the pore wall become equal, hence vanishing the radial potential gradient within the cross-section. On the two sides of this cross-section (axially above and below), the radial component of the local field will have opposite polarities. 
A representative case of this is when $V_0=2V_G$ in the thin EDL limit where the potential on the axis varies independent of the EDL (Fig.~\ref{fig:1D_zeta_gate_vs_V0}(a)). As follows from Eq.~\eqref{eq:zeta-dirichlet-explicit2}, the equivalent zeta potential vanishes at the pore center ($z=l/2$), while taking opposite signs on the two sides of it.

The asymptotic model shows excellent agreement with numerical results in the thin EDL limit. Deviations near the pore ends for finite EDL thickness (Fig.~\ref{fig:1D_zeta_gate_vs_V0}(b)) are attributed to the neglected reservoir-induced effects, such as the overspilling of charge cloud \citep{sherwood2014electroosmosis}, along with the infinitely small buffer region ($\lambda\approx 0$) in the asymptotic analysis. In the limit $\lambda \! \rightarrow \!0$, Eq.~\eqref{eq:bnseries} reduces to a simpler form, shown in Eq.~\eqref{eq:bn} of Appendix \ref{sec:append_zeta_finiteEDL}.
Further, for a long pore, $\kappa l \gg 1$, the local equivalent zeta potential (Eq.~\eqref{eq:zeta-dirichlet-explicit}) evaluated in the bulk of the pore close to the center yields a simplified expression similar to the thin EDL limit, 
\begin{align}
    \zeta_{\mathrm{eq}}(z=l/2) \approx V_G - \frac{V_0}{2},
    \label{eq:zeta_eq_longpore_finiteEDL}
\end{align}
as detailed in Appendix~\ref{sec:append_zeta_finiteEDL}. 
This scaling along with the complete results shown in Fig. \ref{fig:1D_zeta_gate_vs_V0} demonstrates that the axial variation of $\zeta_{\mathrm{eq}}$ is governed primarily by the voltage ratio $V_0/V_G$, and is largely independent of electrolyte concentration. Hence, the dominant mechanism is geometric electrostatic coupling among the radial and axial surfaces rather than bulk ion concentration dependent diffuse-layer thickness effects.

\subsection{Axial asymmetry parameter}
The condition $\zeta_{\mathrm{eq}}=0$ (indicated by the dashed line in Fig.~\ref{fig:1D_zeta_gate_vs_V0}) marks the emergence of axial antisymmetry. This occurs at $V_0=2V_G$, corresponding to a critical balance between axial and radial electrostatic conditions as explained above.
The axial average of the equivalent zeta potential across the pore from Eq.~\eqref{eq:zeta_av} motivates the definition of an axial asymmetry parameter
\begin{align}
    \alpha = 1 - \frac{V_0}{2V_G},
\end{align}
which quantifies the degree of symmetry breaking in the EDL across the pore. The parameter $\alpha$ is independent of bulk ion concentration and depends solely on the ratio of applied voltages.
A critical transition occurs at $\alpha=0$ (i.e., $V_0/V_G=2$), where the pore exhibits axially antisymmetric EDL structure to exhibit an overall electrokinetic behavior having zero average equivalent zeta potential across the pore. In contrast, $\alpha\!=\!1$ corresponds to a uniformly polarized pore with $\zeta_{\mathrm{eq}}\!=\!V_G$.

Fig.~\ref{fig:2D_zeta_eq} compares the space charge density distribution in an axial cross-section of the pore for two distinct surface electrostatic boundary conditions shown in the two halves of the images. In left half, a uniform fixed surface charge (FSC) density ($\sigma$) is prescribed, with the corresponding zeta potential obtained from Grahame’s relation, $\zeta\! = \!(2k_BT/ze_0)\sinh^{-1}\!\left(\sigma/\sqrt{8 \varepsilon \varepsilon_0 c_0 k_BT} \right) \!=\! -25$ mV. In the right half, a uniform fixed surface potential (FSP) of $V_G \!=\! -25$ mV is imposed directly. 
The comparison highlights the modulation of the EDL structure with the axial asymmetry parameter $\alpha$. The two boundary conditions yield identical EDL structures only at $\alpha \!=\! 1$, whereas significant deviations arise elsewhere. In particular, a pronounced axial variation is observed in the radial extent of the space charge distribution away from the surface. For $\alpha < 1$, the EDL undergoes radial thinning, while for $\alpha > 1$, the thickness of the charge cloud increases, as compared to uniform zeta potential scenario. 

At the critical case $\alpha = 0$, the EDL exhibits a fully antisymmetric structure, characterized by opposite charge polarities in the two halves of the pore with no charge at the crosssection through the center of pore. This transition underscores the role of $\alpha$ as a governing parameter for axial symmetry breaking and electrostatic reorganization within the nanopore.  This transition in electrostatic symmetry considerations delineates regimes of qualitatively distinct emergent transport behavior, modulated by the prescribed transmembrane and gate potentials.
This transition identifies $\alpha$ as the key control parameter governing axial symmetry breaking and delineates regimes of distinct emergent transport behavior set by the applied potentials.

Various transport phenomena are explicitly investigated below, for the cylindrical nanopore geometry discussed in Fig. \ref{fig:2D_zeta_eq}.

\section{Transport Implications: Ion Selectivity and Rectification}\label{sec:Results}

\subsection{Ionic Selectivity}\label{subsec:Ion-selectivity}

In nanopores, confinement at length scales comparable to the EDL thickness leads to preferential transport of counter-ionic species under an applied transmembrane potential, commonly referred to as \textit{ion selectivity} ($S$). It is defined, following \citet{tsutsui2024gate}, as
\begin{equation}
    S= \frac{  |I_c| - |I_a| }{|I_c| + |I_a|},
\end{equation}
where $I_c = \int_0^a z_1 \mathbf{N}_1 \cdot \hat{\mathbf{z}} \,2\pi r \,dr$ and $I_a = \int_0^a z_2 \mathbf{N}_2 \cdot \hat{\mathbf{z}} \,2\pi r \,dr$ denote the cationic and anionic current contributions to the axial current at any circular crosssection of the cylindrical pore, respectively.

\begin{figure*}[]
    \centering
    \includegraphics[width=0.32\linewidth]{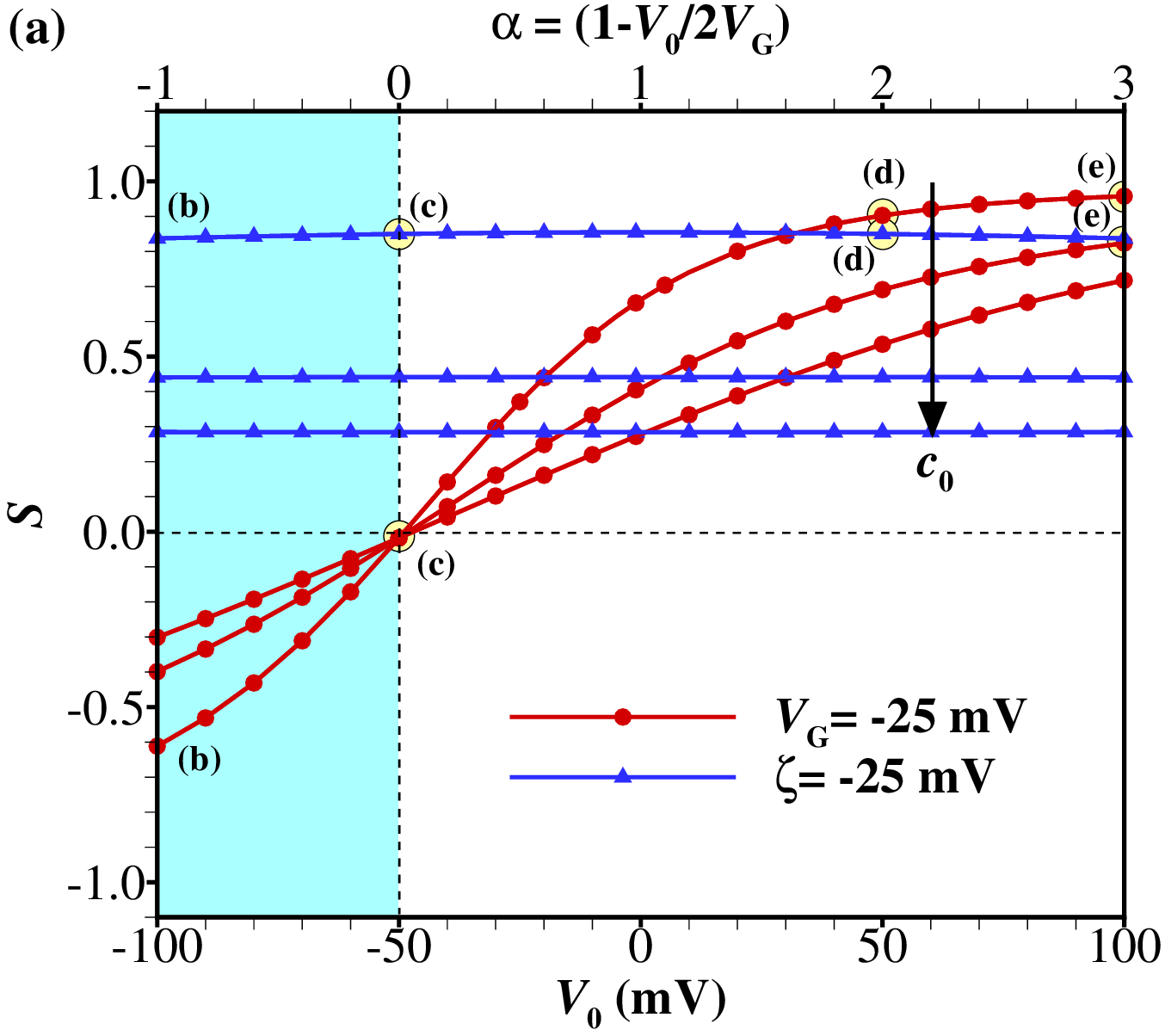}  
    \includegraphics[width=0.60\linewidth]{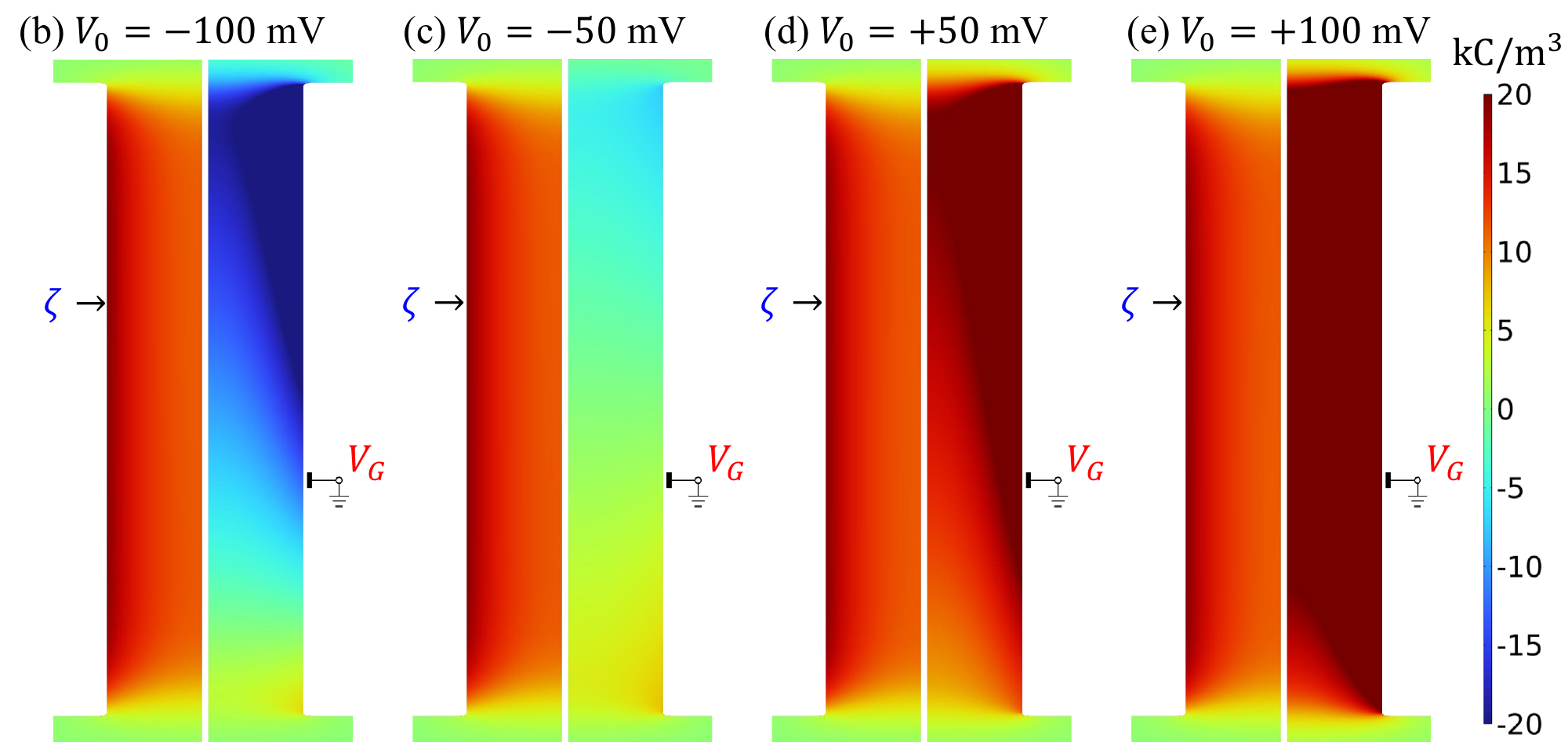} 
    \caption{\textit{Ionic selectivity.} 
(a) Ionic selectivity $(S)$ as a function of transmembrane potential $(V_0)$ for bulk concentrations $c_0=0.04$, $1$, and $10$ mM, corresponding to $\kappa a \approx 1$, $5$, and $16$, respectively. Circular and triangular symbols denote fixed uniform surface potential ($V_G=-25$ mV) and fixed surface charge with uniform zeta potential ($\zeta=-25$ mV) conditions, respectively. 
(b--e) Corresponding space charge density distributions within the nanopore at $c_0=0.04$ mM, comparing fixed zeta potential (left) and fixed surface potential (right) cases for $V_0=-100$, $-50$, $50$, and $100$ mV or asymmetry parameter $\alpha=-1$, $0$, $2$, and $3$, respectively (highlighted in (a)). The radial coordinate is scaled by a factor of 3 for visual clarity.}
    \label{fig:IonSelectivity}
\end{figure*}

Under the fixed surface charge (FSC) condition, ion selectivity originates from nonlinear EDL effects at surface potentials exceeding the thermal voltage. In this regime, excess ion accumulation within the diffuse layer due to counter-ion inflow,
\begin{equation}
\Delta c = (c_+ + c_-) - 2c_0 = 2c_0\left[\cosh\!\left(\frac{e_0\phi}{k_BT}\right) - 1\right],
\end{equation}
enhances the ionic conductivity near the surface \cite{bazant2004diffuse} while depleting the solvent concentration. This results in increased surface conductivity, commonly quantified by the Dukhin number ($Du$) (the ratio of surface to bulk conductivity), and manifests as positive ion selectivity (Fig.~\ref{fig:IonSelectivity}(a)). 
Notably, under FSC conditions, $S$ remains independent of the applied transmembrane potential, consistent with prior theoretical and experimental studies \cite{cervera2006ionic,ramirez2007ion}.

In contrast, fixed surface potential (FSP) in voltage-gated nanopores enable active control of ion selectivity through the interplay of gate potential and transmembrane bias. Experimental studies have demonstrated tunable and even reversible selectivity under such conditions \cite{russell2022gating,tsutsui2024gate,ryzhkov2021switchable,ak2024electrostatic}. Consistently, Fig.~\ref{fig:IonSelectivity}(a) shows that, for $V_G\!=\!-25$ mV, the selectivity strongly depends on the axial asymmetry parameter $\alpha$. Unlike the FSC case, $S$ varies nonlinearly with $\alpha$ and reverses sign below the critical condition $\alpha\!=\!0$, indicating a reversal of ionic current polarity relative to the FSC baseline. The magnitude of selectivity increases and approaches the limiting value $S=1$ where counter-ions are responsible for the whole ionic current for $\alpha>1$, particularly at low bulk concentrations.

This behavior arises from the axially non-uniform EDL structure induced under FSP conditions. For $\alpha<0$, regions with polarity opposite to $V_G$ dominate, leading to reversed selectivity. This is evident from the spatial charge density distribution at low concentration ($c_0\!=\!0.4$ mM) shown in Fig.~\ref{fig:IonSelectivity}(b) for $\alpha\!=\!-1$. At $\alpha=0$, the antisymmetric distribution of $\zeta_{\mathrm{eq}}$ along the pore results in equal and opposite contributions to ionic transport, yielding $S=0$, Fig.~\ref{fig:IonSelectivity}(c) . More generally, the distributions in Fig.~\ref{fig:IonSelectivity}(d, e) illustrate how axial EDL nonuniformity governs the observed selectivity trends for $\alpha>0$.

Also, as the bulk concentration increases, the Debye length decreases, weakening excess EDL conductivity effect despite higher absolute ion densities, leading to reduced selectivity. This phenomenon originates from the spatial variation of driving axial electric field along both axial and radial directions. This can be understood asymptotically for a long pore $\kappa l \gg 1$, where the expression for the axial electric field Eq.~\eqref{eq:axial_Ez} gets rid of higher Fourier modes and reduces to
\begin{align}
    E_z = -\frac{V_0}{l} \left(1 - \frac{I_0(\kappa r)}{I_0(\kappa a)} \right)
    \label{eq:Ez_finiteEDL_longPore}
\end{align}
as discussed in Appendix \ref{sec:append_zeta_finiteEDL}. Here the second term represents a radially varying correction arising from the EDL structure. This expression indicates the decreasing trend of axial field magnitude closer to the surface with prescribed uniform potential. Thus, ion selectivity in FSP is governed by the nonuniform 2D dimensional structure of the EDL induced by electrostatic coupling between $V_0$ and $V_G$, fundamentally differentiating it from FSC systems.

\subsection{Ionic current rectification (ICR)}\label{subsec:ICR}

\begin{figure}[]
    \centering
    \includegraphics[width=0.49\linewidth]{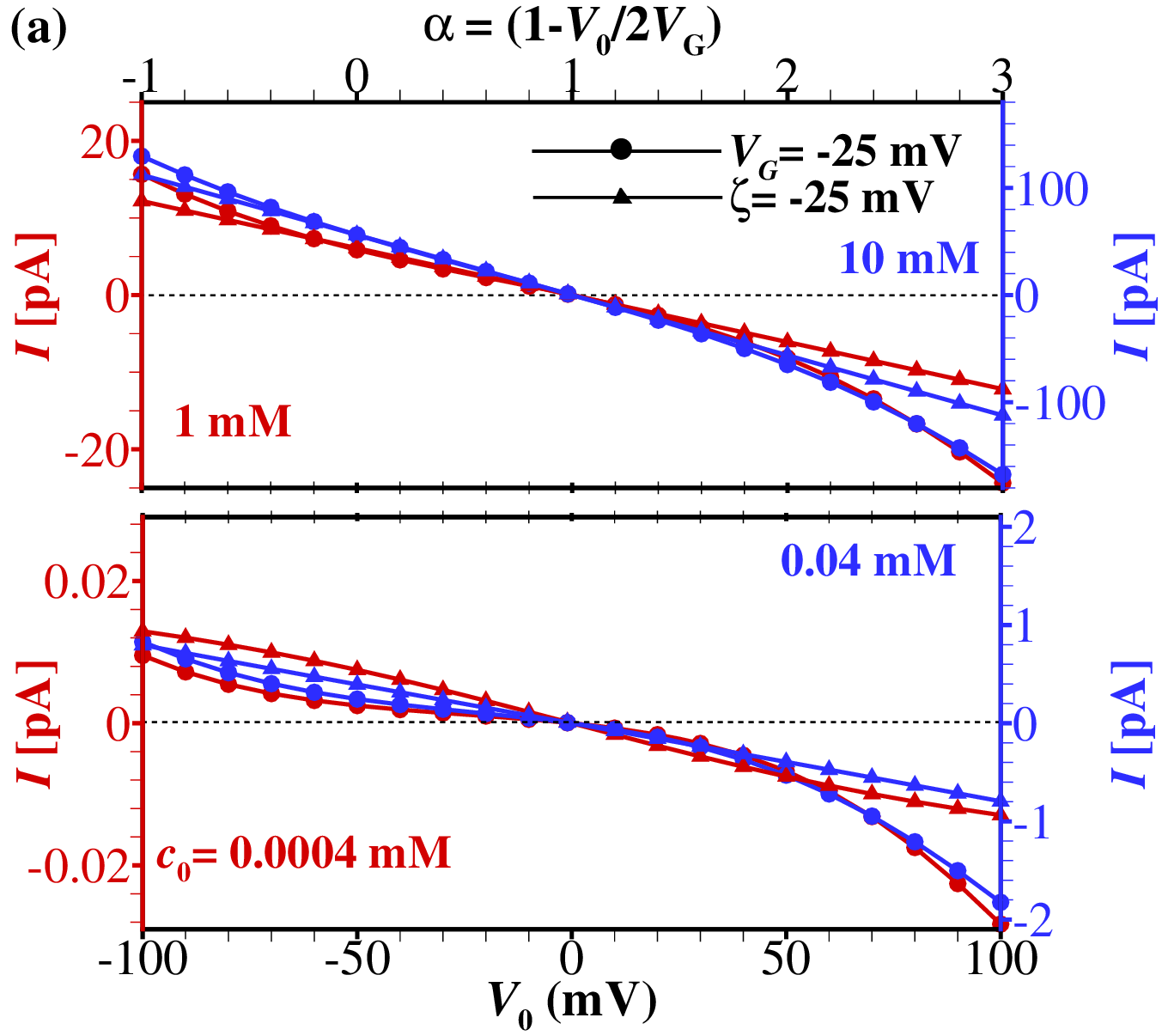} 
    \includegraphics[width=0.49\linewidth]{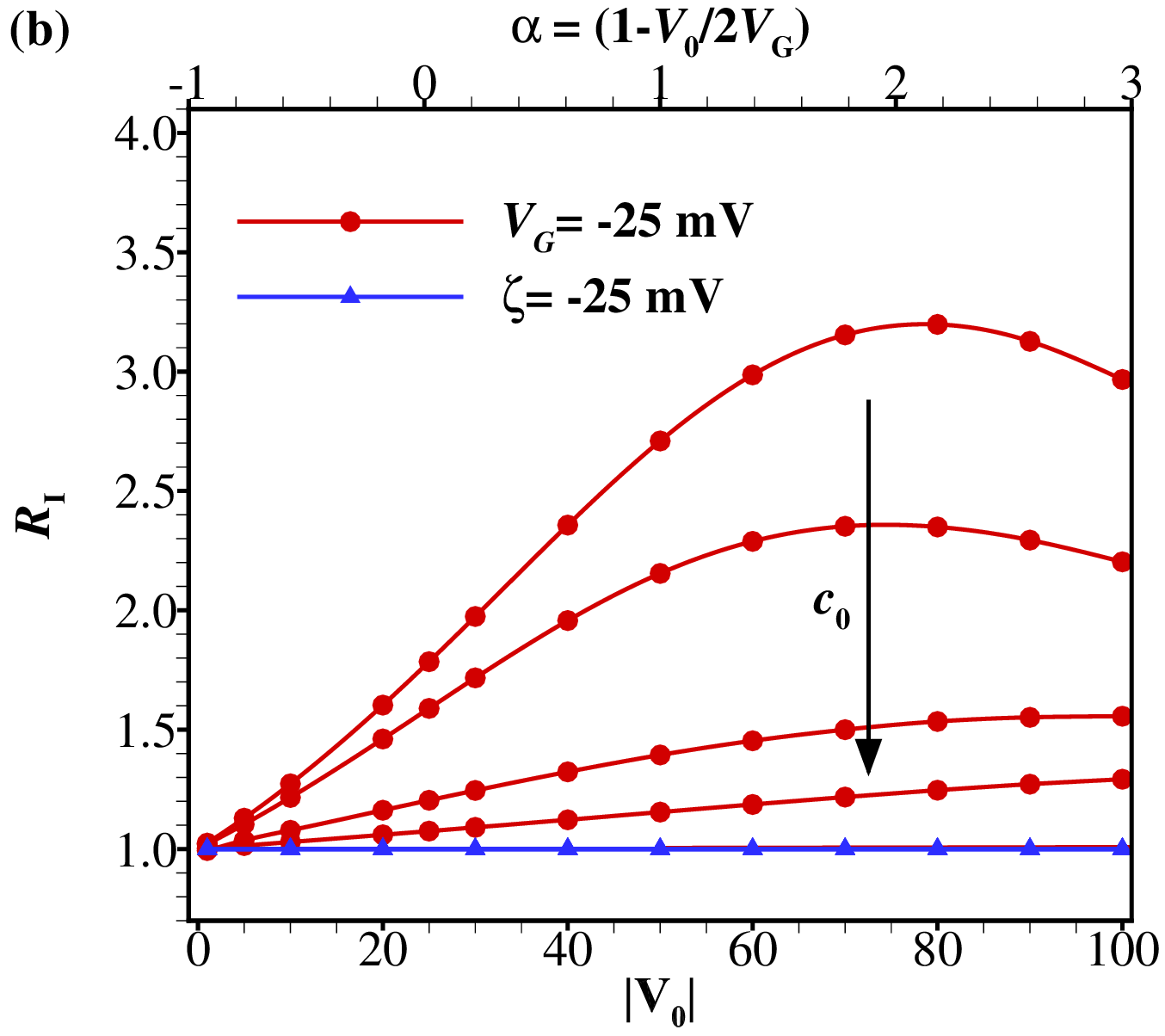} 
    \caption{\textit{Ionic current rectification.} 
(a) Ionic current $I$ (pA) as a function of transmembrane potential $V_0$ (mV) for fixed surface potential ($V_G=-25$ mV, red circles) and fixed surface charge (of $\zeta=-25$ mV, blue triangles) conditions. Results are shown for bulk concentrations $c_0=0.0004$, $0.04$ mM (bottom panel) and $1$, $10$ mM (top panel), corresponding to $\kappa a \approx 0.1$, $1$, $5$, and $16$, respectively. (b) Corresponding rectification factor $R_I$ (Eq.~\eqref{eq:ICR_defin}) as a function of $|V_0|$ for the two boundary conditions.}
\label{fig:ICR}
\end{figure}

While the selectivity $S$ characterizes the relative ionic contributions, the net axial ionic current magnitude $I\! =\! I_c + I_a$ governs the overall transport through the pore. Figure~\ref{fig:ICR}(a) shows the variation of $I$ with transmembrane potential $V_0$ for bulk concentrations $c_0\!=\!4\times10^{-4}$, $0.04$, $1$, and $10$ mM. As expected, the current magnitude increases with increasing $c_0$ due to enhanced conductivity.
Under fixed surface charge (FSC) conditions, the current varies nearly linearly with $V_0$, exhibiting symmetric behavior with respect to polarity. 
In contrast, fixed surface potential (FSP) conditions display pronounced nonlinear and asymmetric current--voltage characteristics at larger $|V_0|$, similar to that reported by \citet{chen2025electrostatically}. 

This asymmetry is quantified using the rectification ratio \citep{zhang2024modulation}
\begin{equation}
    R_{I}= -\frac{I(+|V_0|)}{I(-|V_0|)} \label{eq:ICR_defin}
\end{equation}
and shown in Fig. \ref{fig:ICR}(b). While FSC shows negligible rectification ($R_I\! \approx \!1$), the FSP case exhibits significant rectification, with $R_I$ reaching values up to $\sim 3$. 
Notably, the rectification exhibits a non-monotonic dependence on $|V_0|$, with a distinct peak whose location shifts with bulk concentration .
Unlike ion selectivity, the total ionic current remains symmetric with respect to ion polarity. Consequently, the antisymmetric EDL configuration at $\alpha\!=\!0$ does not lead to qualitative transition in current direction, and the current continues to increase with the applied potential gradient, similar to the FSC case. The rectification therefore arises not from conductivity variation, but from an inherent axial transport asymmetry within the pore.

The origin of this rectification can be traced to the two-dimensional nonuniformity of the EDL structure under FSP conditions. As shown in Fig.~\ref{fig:2D_zeta_eq} and Fig.~\ref{fig:IonSelectivity}(b--e), the space charge distribution varies both axially and radially as indicated by asymptotic analysis in Eq.~\eqref{eq:space_chrarge_analytical}. This induces an additional axial electric field component within the EDL, which modifies the effective driving field experienced by the ions. As a result, the transport symmetry of ionic current response with respect to the polarity of $V_0$ is broken, leading to ionic current rectification.

This mechanism can be understood from the analytical expression for the axial electric field, Eq.~\eqref{eq:axial_Ez} , within a finite length of the pore ($\kappa l \! \not\gg \! 1$). The second term of this expression involves an internal axial field within the EDL due to the interaction of axial and radial prescribed potentials. This term effectively acts as a bias to the externally applied field, altering the axial driving force for ion transport. In the thin EDL limit, this correction vanishes, restoring symmetric current--voltage behavior and eliminating rectification. Accordingly, the rectification decreases with increasing electrolyte concentration.
Especially at low concentrations, where the EDL thickness becomes comparable to the pore dimensions ($\kappa l\! \sim \!1$) (Fig.~\ref{fig:ICR}(b)), there is a competition between the applied field and the EDL-induced bias. As $|V_0|$ increases, the external field dominates, reducing the relative influence of the EDL contribution. This results in a non-monotonic variation in rectification magnitude, with a peak in $R_I$ occurring at intermediate values of $\alpha$ (typically around $\alpha\! \approx \!2$ at low concentrations).

\subsection{Electroosmotic flow (EOF) rectification}\label{subsec:EOF}
\begin{figure*}[]
    \centering
    \includegraphics[width=0.26\linewidth]{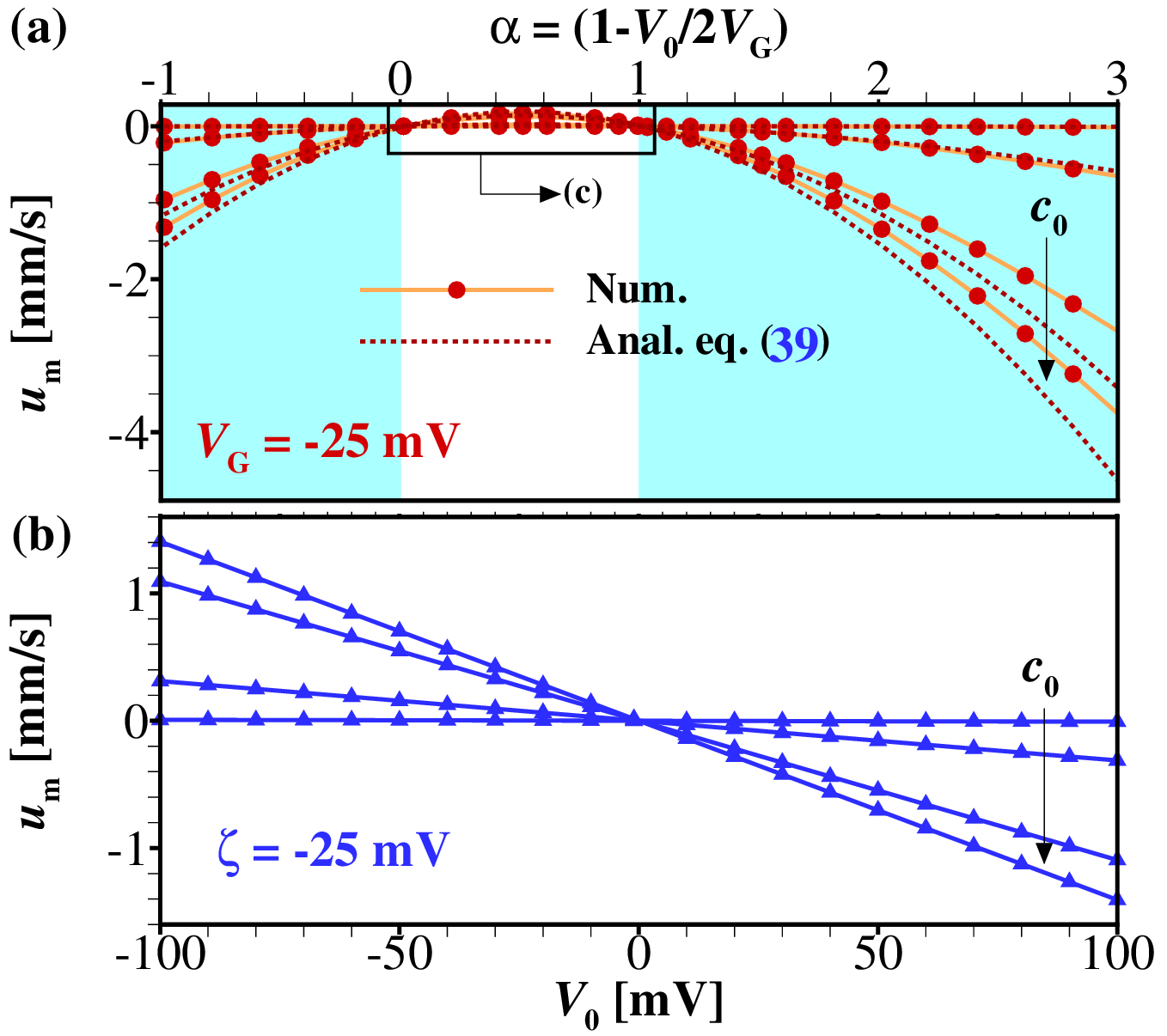}  
    \includegraphics[width=0.26\linewidth]{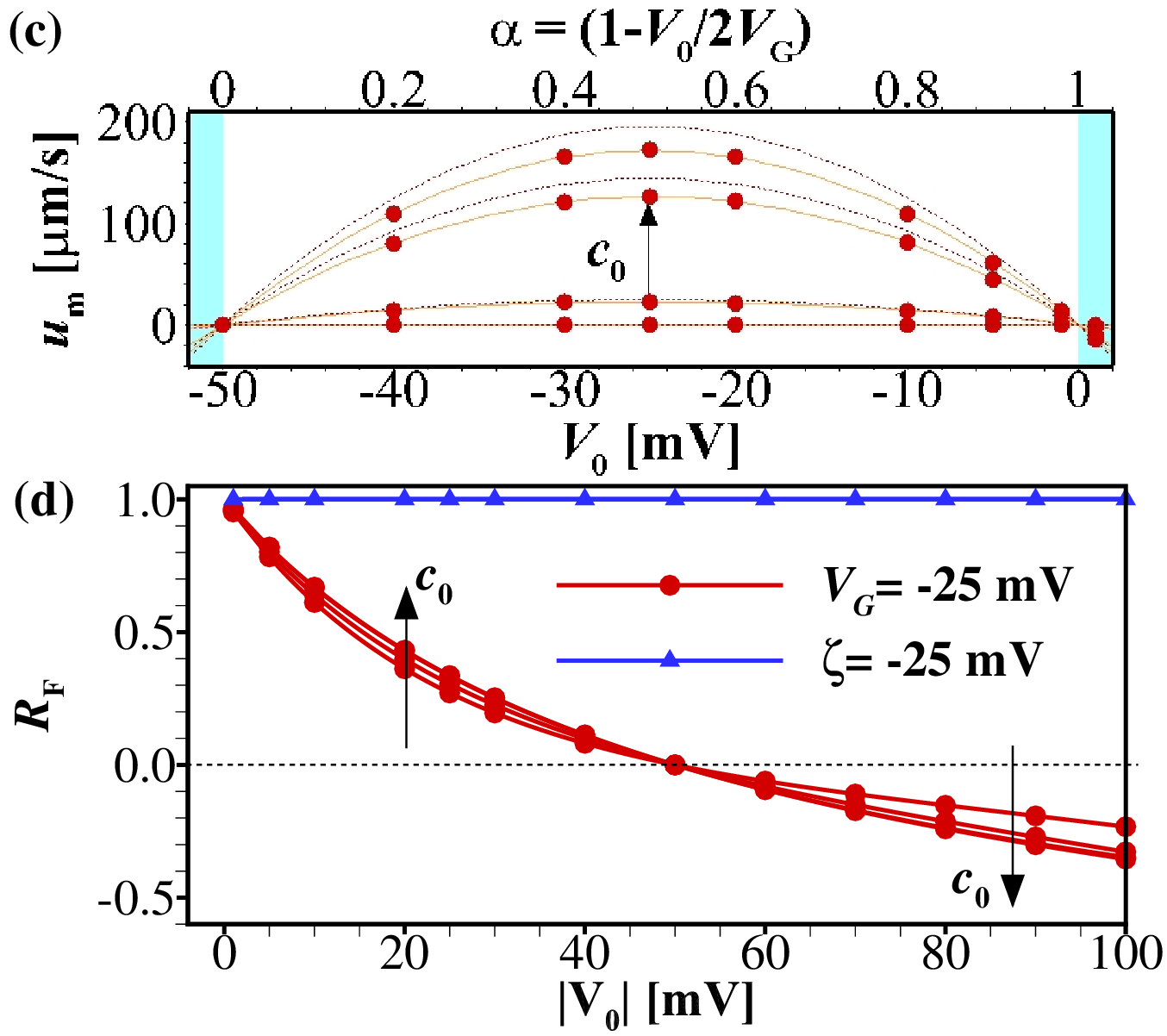}
\includegraphics[width=0.42\linewidth]{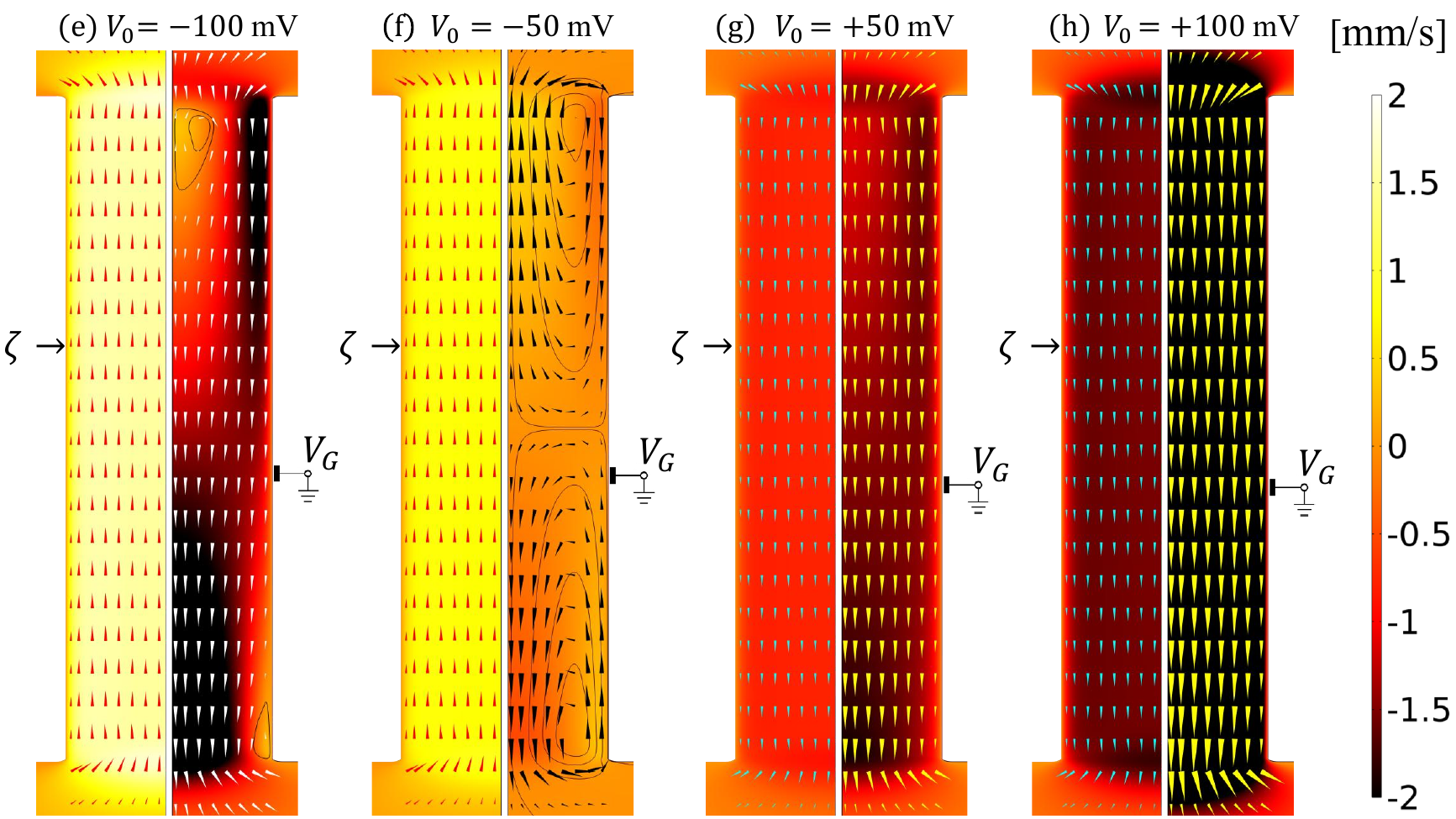}  
    \caption{\textit{EOF rectification:} (a,b) Average velocity ($u_{av}$) verses applied external potential $V_0$ at two different scenarios of (a) $\zeta\!=\!-25$ mV and (b) gate potential $V_G\!=\!-25$ mV on the nanochannel wall. Here, four different values of the bulk concentration is considered as $c_0\!=\!0.0004, 0.04, 1, 10$ mM, which corresponds to $\kappa a \! \approx \!$ 0.1, 1, 5, 16, respectively. (d) Corresponding rectification factor $R_{f,EOF}$ as a function of $|V_0|$ for the two different cases. Lines with circular symbols and delta symbols represents the results for a fixed gate potential ($V_G\!=\!-25$ mV) and fixed zeta potential ($\zeta\!=\!-25$ mV) case, respectively. (e-f) At $c_0\!=\!10$ mM, the axial velocity (background) and corresponding velocity vectors are shown for two different cases of fixed surface potential (right) and fixed zeta potential (left) at the nanopore wall for $V_0\!=\!$ -100 mV(e), -50 mV(f), +50 mV(g), and +100 mV(h).}
    \label{fig:Velocity}
\end{figure*}

Electroosmotic flow (EOF) in the nanopore is driven by the axial electric force acting predominantly on the EDL charge, subject to the no-slip boundary condition at the pore walls. Since both the electric field and space charge density exhibit strong spatial variations due to the coupling between transmembrane and gate potentials, the resulting flow inherits a similarly rich behavior.
The volumetric flow of the incompressible fluid through the pore is quantified using the magnitude of cross-sectional mean axial velocity (along the positive z-axis direction) $u_m$ from Eq.~\eqref{eq:velocity_av}, which is uniform throughout the length of the pore. The flow evaluated using both numerical simulations (via cross-sectional integration) and the asymptotic analysis are plotted in Fig.~\ref{fig:Velocity}. It shows the variation of $u_{\mathrm{m}}$ with $V_0$ for different bulk concentrations under fixed surface potential conditions.

Under fixed surface charge (FSC) conditions with a negatively charged pore, the flow exhibits a linear and antisymmetric dependence on $V_0$ (Fig.~\ref{fig:Velocity}(b)). Reversing the transmembrane potential reverses the flow direction while preserving its magnitude. The flow remains aligned with the applied axial electric field, and its magnitude increases with bulk concentration due to enhanced charge density within the EDL.

In contrast, fixed surface potential (FSP) conditions lead to three distinct flow regimes (Fig.~\ref{fig:Velocity}(a,c)). For $\alpha\!>\!1$, the flow is aligned with the axial electric field, whereas for $\alpha\!<\!0$, it opposes the applied field despite the same sign of $V_G$. In this regime, reversing $V_0$ polarity does not reverse the flow direction but only changes its magnitude. In the intermediate regime ($0<\alpha<1$), the flow is in the opposite direction to the other regimes, while its direction reverses with $V_0$.

This leads to an electroosmotic flow rectification, quantified by
\begin{equation}
    R_F = -\frac{u_{\mathrm{m}}(-|V_0|)}{u_{\mathrm{m}}(+|V_0|)}.
\end{equation}
Figure~\ref{fig:Velocity}(d) shows that $R_F$ exhibits both positive and negative rectification. Positive rectification ($R_F>0$) occurs in the intermediate regime with $|V_0/V_G|<2$, where flow reverses with $V_0$; while a peculiar negative rectification ($R_F<0$) arises in the other two regimes for $|V_0/V_G|>2$, where the flow direction remains unchanged upon reversing $V_0$, as illustrated by comparing the velocity fields in Fig.~\ref{fig:Velocity}(e,h).

The origin of this behavior lies in the nonlinear electrostatic forcing induced by the nonuniform EDL charge density as shown in figure \ref{fig:IonSelectivity}(b-e). In addition to that, similar to ionic current rectification, the axial field acquires a spatially varying correction due to the EDL structure, which biases the effective driving force. As a combined effect, it results in the velocity deviation from a linear response at larger $|V_0|$, leading to rectification.

This mechanism is captured by the asymptotic expression for the mean velocity (Eq.~\eqref{eq:velocity_av}), which depends on the product $E_z(r,z)\,\zeta_{\mathrm{eq}}(z)$. The thin-EDL asymptotic predictions (dashed lines) agree closely with the numerical results (solid lines), with good accuracy retained even for moderately thick EDLs (Fig.~\ref{fig:Velocity}(b,c)). Consequently, $u_m$ exhibits a quadratic dependence on $V_0$, resulting in two critical points of zero net flow at $V_0\!=\!0$ and $V_0\!=\!2V_G$ (Fig.~\ref{fig:Velocity}(c)). This quadratic behavior directly explains the emergence of negative rectification for $|V_0/V_G|\!>\!2$.

Although $u_{\mathrm{m}}\!=\!0$ at both $\alpha\!=\!1$ and $\alpha\!=\!0$, the underlying flow structures are fundamentally different. At $\alpha\!=\!1$, the EDL is axially uniform, leading to identically zero velocity throughout the pore. In contrast, at $\alpha\!=\!0$, the antisymmetric EDL generates opposing flows in the two halves of the pore, resulting in internal recirculation and a vortex-pair formation in each half-crosssection, as shown in Fig.~\ref{fig:Velocity}(f). 

Overall, the coupling between nonuniform axial electric field ($E_z(r,z)$ from Eq.~\eqref{eq:axial_Ez}) and spatial EDL charge distribution ($\rho_e(r,z)$ from Eq.~\eqref{eq:space_chrarge_analytical}) produces a nonlinear and asymmetric electrostatic forcing, giving rise to unique EOF modulation and peculiar rectification phenomena that are absent in classical FSC systems.

\section{Discussion}
Ion selectivity, ionic current rectification (ICR), and electroosmotic flow rectification (EFR) in nanopores have traditionally been studied as distinct phenomena, often realized through different design strategies to induce symmetry breaking. In contrast, the present work demonstrates that, under fixed surface potential (FSP) conditions, these seemingly disparate responses emerge from a single underlying mechanism: axial nonuniformity of the EDL induced by electrostatic coupling between the externally prescribed transmembrane ($V_0$) and gate ($V_G$) potentials.
Within this framework, ion selectivity reflects the axially nonuniform EDL structure and charge polarity (through $\zeta_{\mathrm{eq}}(z)$ Eq.~\eqref{eq:zeta-dirichlet-explicit}), ionic current rectification arises from the internal axial field bias (in Eq.~\eqref{eq:axial_Ez}) resulting from the nonuniform distribution of space charge density that breaks voltage symmetry, and flow rectification emerges from the nonlinear coupling between this biased field and the spatially varying EDL charge ($E_z(r,z) \rho_e(r,z)$ from Eq.~\eqref{eq:axial_Ez}, \ref{eq:space_chrarge_analytical}). 

The same mechanism also leads to nontrivial flow structures, including the emergence of vortices as velocity and pressure fields reorganize to satisfy flux conservation. Tokens of such nanoscale vortical flows were indicated to arise from geometry-driven asymmetries of conical nanopores in the literature \cite{zeng2019rectification,yao2020induced}. However, the present results explicitly show pronounced engineered vortical structures with their mechanistic origins.  

Remarkably, all three transport responses are governed by a single asymmetry parameter $\alpha$, with consistent transitions in selectivity, current symmetry, and flow direction occurring at critical values of $\alpha$. This establishes coupling of axial and radial electric fields as the fundamental control parameter for electrokinetic transport in voltage-gated nanopores.
The present analysis not only provides a mechanistic foundation for recent experimental observations of tunable selectivity and rectification in gated nanopores \citep{tsutsui2024gate,ak2024electrostatic,chen2025electrostatically}, but also isolates the essential physics by operating within moderate surface potentials ($V_G \sim$ thermal voltage). This reveals that the observed functionalities do not rely on extreme driving conditions, but instead arise from a deeper electrostatic coupling that can be systematically controlled.


Across nanopore systems, symmetry breaking underlying transport phenomena is typically quantified using method-specific asymmetry parameters, such as geometric taper ratios in conical pores \citep{siwy2006ion}, surface patterning fractions in bipolar nanopores \citep{zhang2024modulation}, or gradients in salinity \citep{cao2012concentration,deng2014effect} and viscosity \citep{qiu2018abnormal}. While these parameters capture qualitative regimes and quantitative trends within their respective contexts, they remain intrinsically tied to the specific implementation of asymmetry.
In contrast, the present work motivates a direct, method-agnostic description of asymmetry through its underlying physical origin, the axial nonuniformity of the EDL, quantified via the length-averaged equivalent zeta potential (Eq.~\eqref{eq:zeta_av}). In the present electrostatic framework, this nonuniformity is governed by the interplay between gate and transmembrane potentials, whereas in other approaches it would arise from geometric, material, or operational variations.
This motivates a generalized asymmetry index $\overline{\alpha}$ defined as $\langle \zeta_{\mathrm{eq}} \rangle = \overline{\alpha}\, V_0$, which captures the effective electrostatic symmetry breaking independent of its method. 

In practical nanopore systems, multiple sources of asymmetry often coexist (for example conical voltage-gated nanopores with inherent charge from \citet{hong2024rectification}). Although mechanistically separated in the present study, the surface charge arising from interfacial chemistry, fabrication-induced heterogeneity, and externally applied fields are inherently coupled, and their combined influence governs the overall transport response. The present framework provides a systematic way to interpret such complex scenarios, with the asymmetry parameter serving as a unifying descriptor across different symmetry-breaking methods. This also suggests that electrostatic gating can be integrated with geometric or chemical design to achieve enhanced, tunable as well as actively controllable transport functionalities.

The present results indicate that simplified boundary representations can capture the essential transport behavior of complex surfaces in advanced 2D materials, although the ion-specific surface chemistry would improve the quantitative predictions. 
Similarly, extending the present analysis to strongly nonlinear regimes and solvent pH or molecular networks influenced EDL formation remains an important direction for future theoretical and analytical work, particularly for bridging with operating conditions in emerging voltage-gated nanopore technologies. Ongoing efforts are focused on extending the present continuum framework to incorporate molecular-scale effects, including dipolar solvent molecular interactions \citep{srinivasula2026dipolar}, explicit solvent molecular network dynamics \citep{srinivasula2026electrohydrodynamic}, and ion-solvent interactions \citep{pandey2021impact} among other possible enrichments, to improve predictive accuracy in nanopore transport. While the present steady-state analysis captures the essential electrokinetic behavior, time-dependent simulations will be important for resolving transient phenomena, particularly in the transport of nanoparticles and biomolecules (e.g., DNA) in voltage-gated nanopores.

From an applications perspective, the mechanistic understanding improves the ability to dynamically control asymmetry through gate potential and opens new avenues for physics inspired integration with data-driven approaches in nanofluidic systems, in particular toward programmable nanopores for biosensing, energy harvesting, and nanofluidic logic.
Beyond existing applications, the present results from the FSP nanopores enable unique such as, ion segregation and spatial focusing at specific axial locations, which may be exploited for selective ion capture, desalination, and localized electrochemical reactions within the pore \cite{tsutsui2025transmembrane}. Furthermore, the emergence of internally generated vortical structures and negative flow rectification suggests opportunities for enhanced nanoscale mixing and stronger AC electroosmotic pumping without relying on geometric asymmetry, in contrast to conventional conical nanopores \cite{wu2016alternating}, respectively.

\section{Conclusions}\label{sec:conclusions}

In contrast to classical fixed surface charge (FSC) nanopores, where transport is governed primarily by radial EDL structure, the present fixed surface potential (FSP) system reveals a strong coupling between axial and radial electrostatics. This coupling induces axial EDL nonuniformity, explaining fundamentally different and richer transport behavior from the prior experimental observations .
At surface (gate) potentials on the order of the thermal voltage, the analysis captures key experimental trends in ion selectivity and ionic current rectification (ICR), while additionally predicting distinct regimes of electroosmotic flow rectification (EFR), including both positive and negative responses. These diverse transport phenomena are consistently described using a single mechanistic asymmetry parameter $\alpha$, which quantifies the net effect of the axial EDL nonuniformity across the pore.

A critical condition arises at $\alpha=0$, where the pore exhibits perfect axial antisymmetry, where ion selectivity and EOF rectification undergo sign reversal, marking transitions between qualitatively distinct transport regimes. More generally, the axial nonuniformity of the EDL generates internal electric field and force biases, giving rise to nonlinear transport responses, current rectification, and the emergence of vortical flow structures within the nanopore, respectively.

Overall, this work establishes axial electrostatic symmetry breaking as a unifying mechanism underlying selectivity, current rectification, and flow modulation in voltage-gated nanopores. Also, by framing asymmetry in terms of EDL structure rather than specific implementation, it provides a mechanistic foundation for systematically designing and controlling advanced, programmable transport functionalities in nanofluidic systems.

\printcredits

\textbf{Declarations}
DP gratefully acknowledges financial support from the Indian Institute of Technology Mandi under Grant No. IITM/SG/2025/06-169.

Authors express no conflicts of interest.

Simulation codes may be available from the authors upon reasonable request. All other data are included within the manuscript.

\appendix

\section{Numerical modeling details}\label{App:Numerical_details}

\begin{figure}
    \centering
    \includegraphics[width=0.9\linewidth]{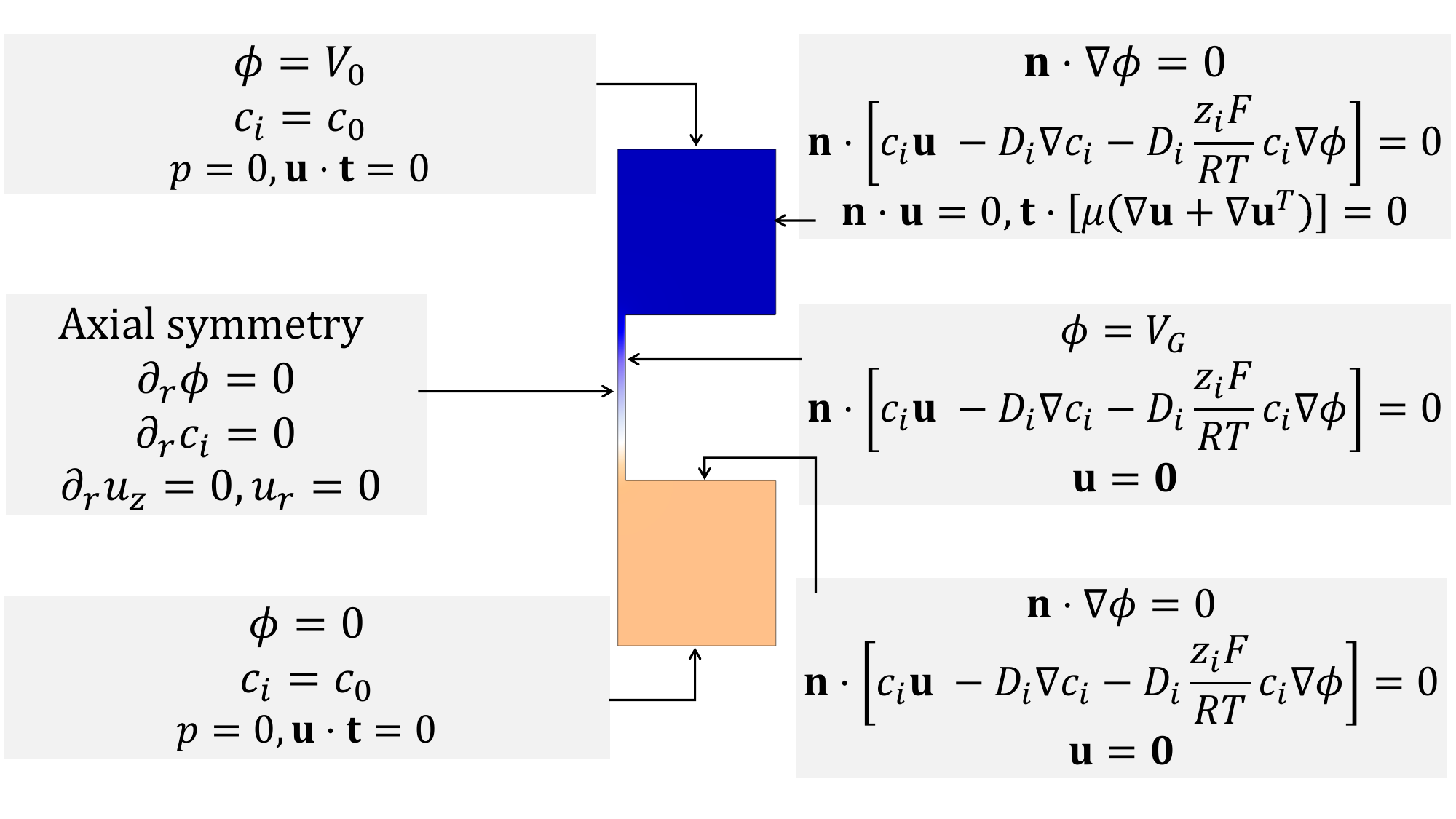}
    \caption{Boundary conditions at each boundary of the computational domain shown in a representative crosssection of the axisymmetric geometry for the fixed surface potential operation. Identical boundary conditions are considered at the lateral and membrane contact boundaries of the top and bottom reservoir.}
    \label{fig:App_BC}
\end{figure}

The governing equations constitute a tightly coupled multiphysics system. The electrostatic potential $\phi$ depends on ionic concentrations through the Poisson equation (Eq.~\eqref{eq:Poisson_phi}); the ionic fluxes depend on $\phi$ and the fluid velocity $\mathbf{u}$ via the Nernst--Planck equations (Eq.~\eqref{eq:NPeq}); and the velocity field $\mathbf{u}$ depends on $\phi$ and ionic distributions through the electric body force in the Stokes equations (Eq.~\eqref{eq:Stokes}). The boundary conditions for the computational domain are summarized in Fig.~\ref{fig:App_BC}.

\begin{figure}
    \centering
    \includegraphics[width=0.48\linewidth]{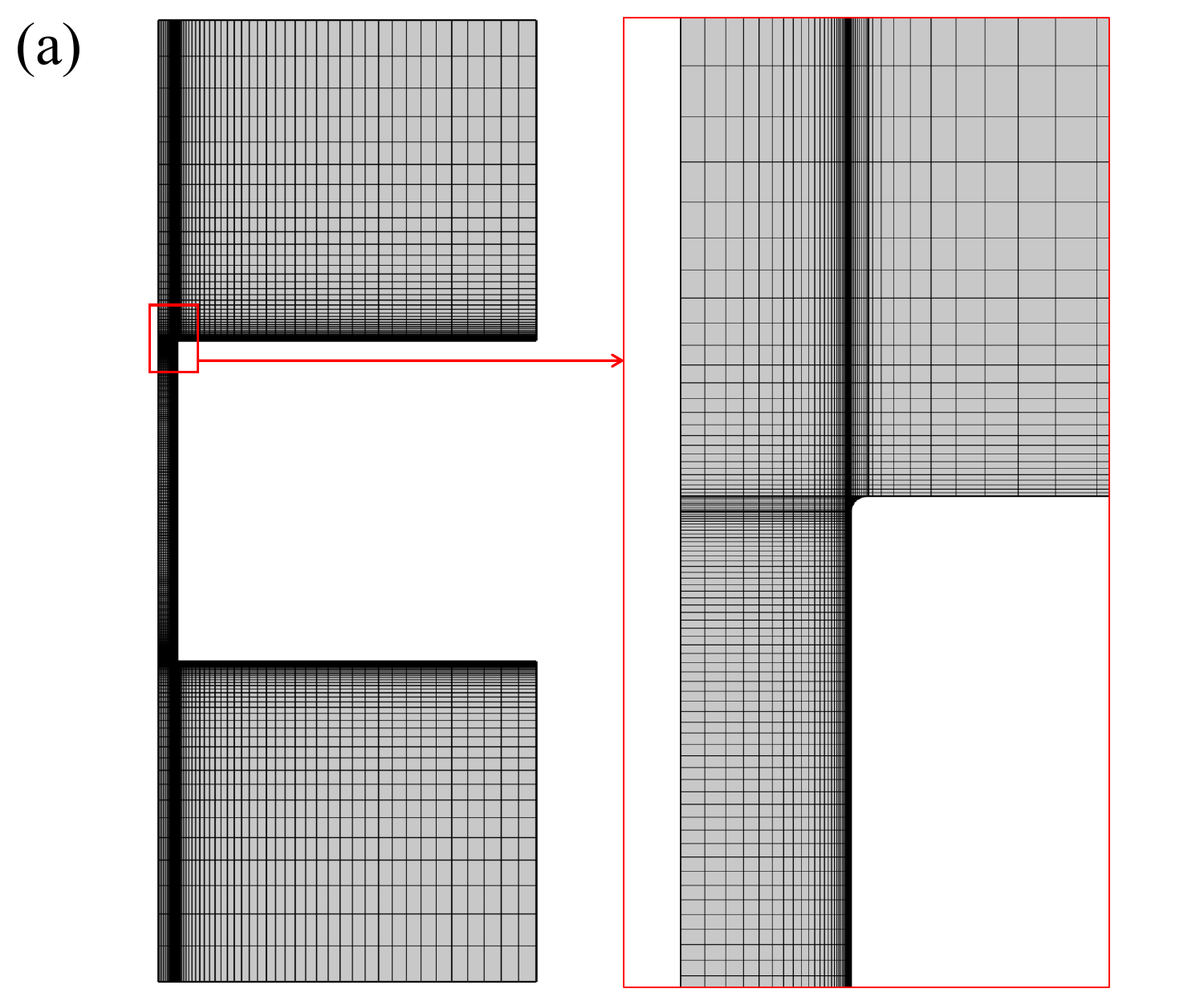}
    \includegraphics[width=0.45\linewidth]{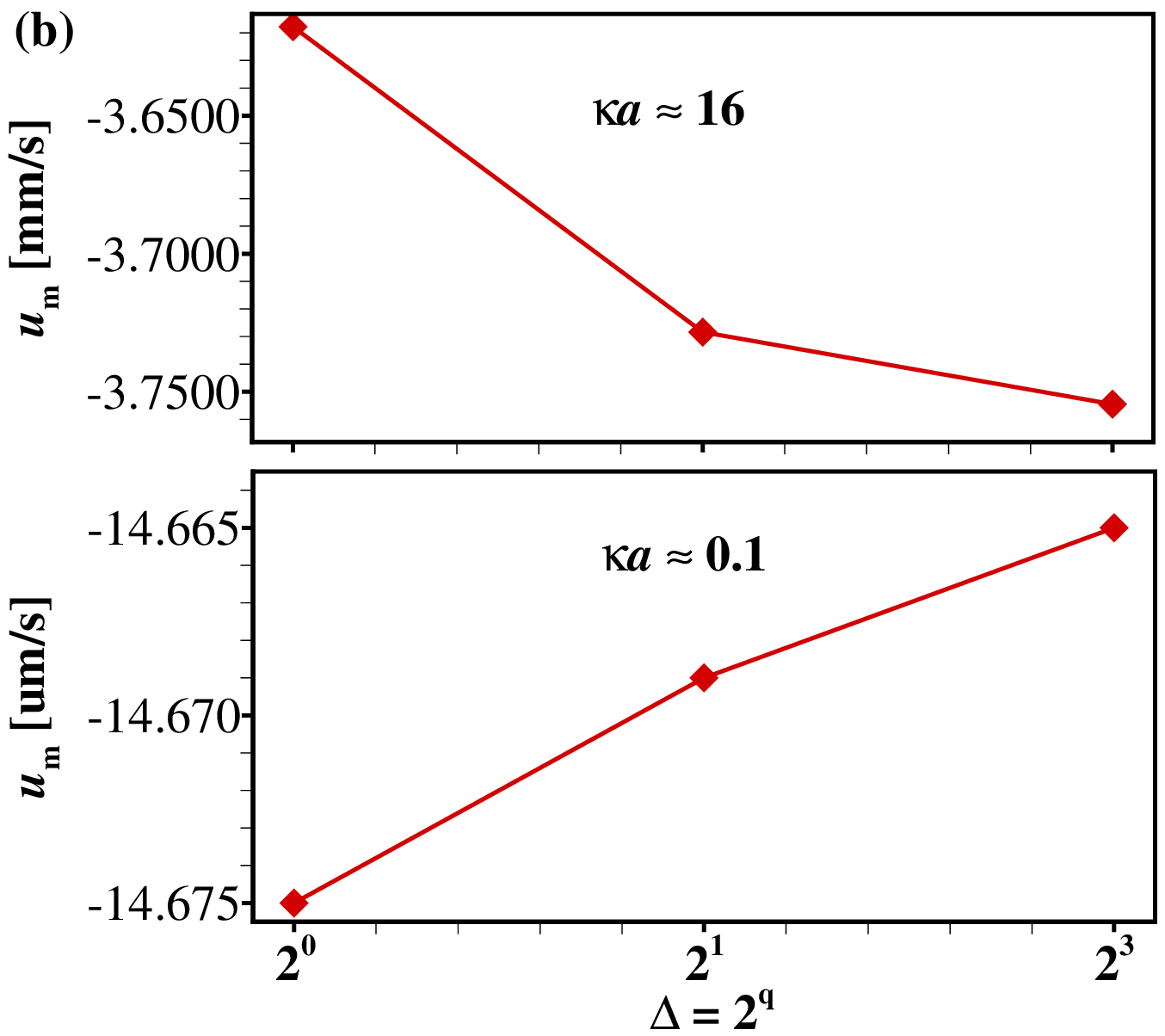}
    \includegraphics[width=0.45\linewidth]{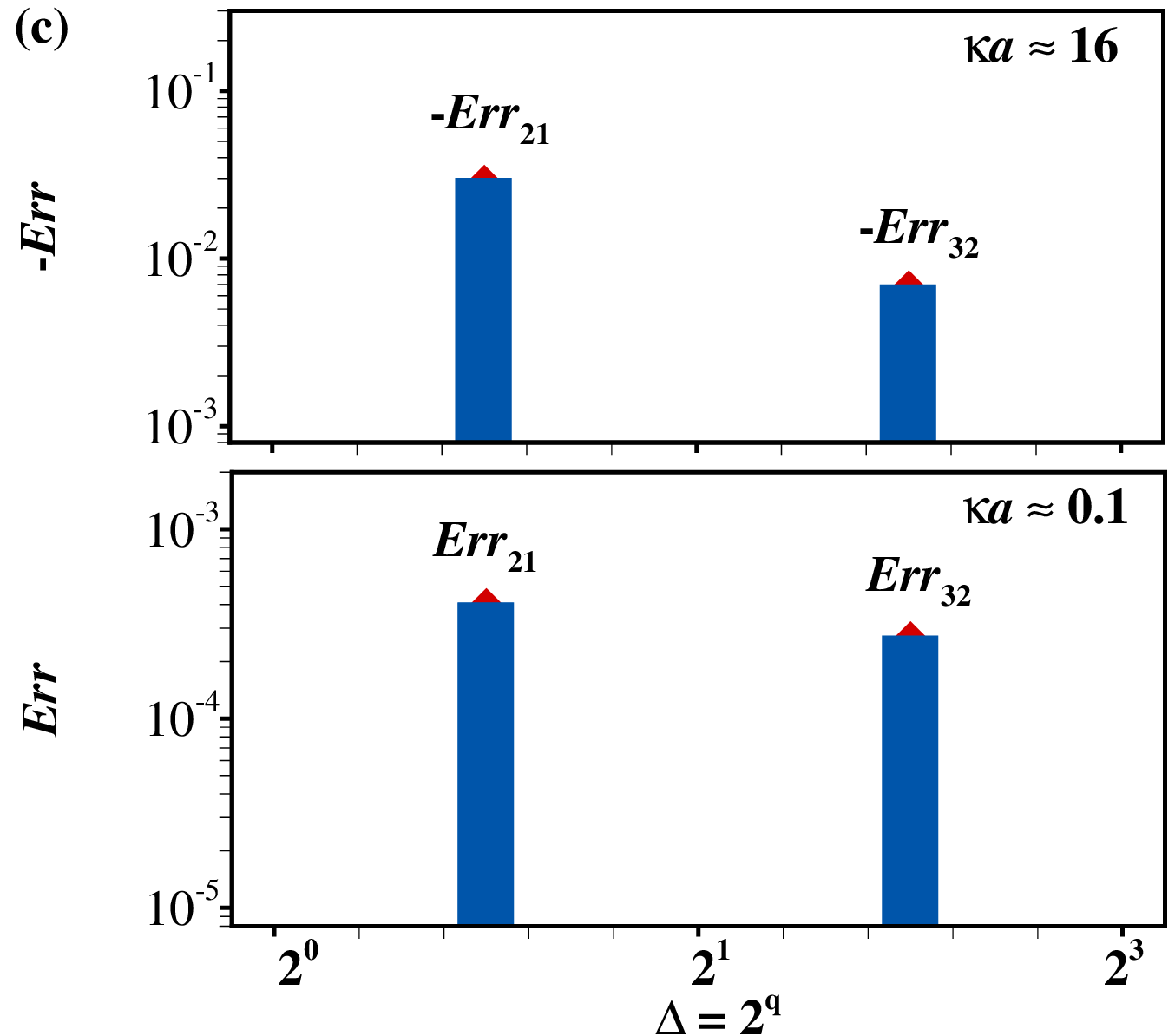}
    \includegraphics[width=0.45\linewidth]{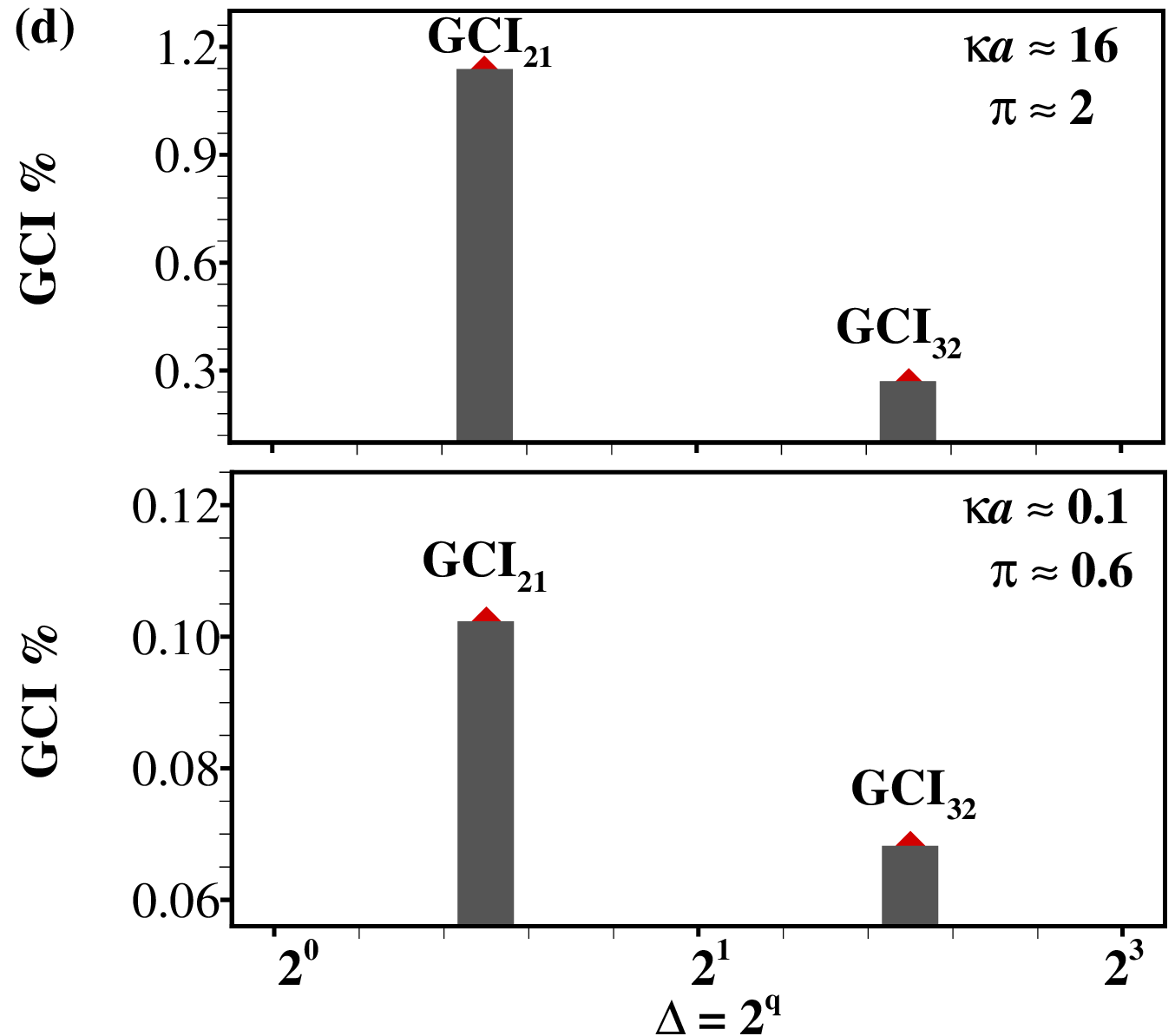}
    \caption{(a) Grid distribution corresponding to $\Delta=2^1$. However, $\Delta=2^2$ is considered for obtaining all results presented in this study. The corner region at the pore-reservoir junction is highlighted for clarity. (b) Average velocity $u_{m}$ for three sets of grid distributions corresponding to $\Delta=2^0,2^1$ and $2^2$ at $\kappa a\approx 0.1$ and 16. The corresponding relative error (c) and grid convergence index (GCI) percentage (d) for the same. Other parameters are kept fixed at $a=50$ nm, $l=1~\mu$m, $V_0=100$ mV, and $V_G=-25$ mV. }
    \label{fig:App_grid}
\end{figure}

The system is discretized using the finite element method (FEM). Quadratic shape functions are employed for the electric potential ($\phi$), ionic concentrations ($c_i$), and velocity ($\mathbf{u}$), while pressure ($p$) is discretized using linear shape functions, ensuring numerical stability. Accurate resolution of the electric double layer (EDL), hydrodynamic boundary layers, and geometric transitions at the pore--reservoir interface requires a nonuniform mesh (Fig.~\ref{fig:App_grid}(a)). Structured rectangular elements are used in the bulk region, while locally refined unstructured triangular elements are employed near boundaries and geometric singularities to balance accuracy and computational cost.

Grid independence is verified using Richardson extrapolation and the grid convergence index (GCI). A refinement parameter $\Delta = 2^q$ is introduced, with $q = 0, 1, 2$ corresponding to coarse, medium, and fine grids, respectively, yielding a refinement ratio of 2. The grid density in both radial and axial directions scales with $\Delta$. The medium grid ($\Delta=2^1$) is illustrated in Fig.~\ref{fig:App_grid}(a). The cross-sectionally averaged velocity ($u_{m}$) is used as the representative quantity for convergence assessment.

Since EDL thickness influences numerical resolution, convergence is evaluated for both thick and thin EDL limits, corresponding to $\kappa a \!\approx \! 0.1$ and $16$. The relative errors between successive grid levels are defined as
\begin{align}
    Err_{21} &= \frac{u_{m}|_{q=2} - u_{m}|_{q=1}}{u_{m}|_{q=1}}, \\
    Err_{32} &= \frac{u_{m}|_{q=3} - u_{m}|_{q=2}}{u_{m}|_{q=2}}.
\end{align}
The observed solutions exhibit monotonic convergence for both limits (Fig.~\ref{fig:App_grid}(b)). The order of convergence is estimated as
\begin{equation}
    \pi = \frac{\ln|Err_{21}/Err_{32}|}{\ln(2)}.
\end{equation}
Using a safety factor $F_s = 1.25$, the GCI values are computed as
\begin{align}
    \text{GCI}_{21} &= \frac{F_s |Err_{21}|}{2^\pi - 1} \times 100\%, \\
    \text{GCI}_{32} &= \frac{F_s |Err_{32}|}{2^\pi - 1} \times 100\%.
\end{align}

The results (Fig.~\ref{fig:App_grid}(c,d)) show that the GCI remains well-below 1\% for the thick EDL case, and within $\sim 1.1\%$ (coarse-to-medium) and $\sim 0.3\%$ (medium-to-fine) for the thin EDL case. The asymptotic convergence criterion is also satisfied, confirming that the solution approaches the grid-independent limit. Based on this analysis, the finest grid ($\Delta=2^2$) is used for all reported results.

The resulting nonlinear algebraic system is solved using a fully coupled Newton method with a direct PARDISO solver. For representative parameter sets, steady-state solutions are obtained within approximately 10 minutes (thick EDL) and 20 minutes (thin EDL) on a 16-core 4\,GHz processor with 32\,GB RAM.

\section{$\zeta_{eq}$ in a long pore for finite EDL thickness}\label{sec:append_zeta_finiteEDL}



In the limit of vanishingly small buffer regions of surface conditions at the corners, $\lambda \ll 1$,  Eq.~\eqref{eq:bnseries} yields Fourier coefficients of $b_n$ that are independent of the axial coordinate as,
\begin{equation}
b_n = \frac{2}{n\pi} \left[
V_G (1 - (-1)^n) + V_0 (-1)^n
\right].\label{eq:bn}
\end{equation}
We now evaluate the relevant series term of \ref{eq:zeta-dirichlet-explicit} at $z=l/2$ as
\begin{align}
\!\!\!\!\!\sum_{n=1}^{\infty}
\frac{b_n}{I_0(\alpha_n a)}
\sin\!\left(\frac{n\pi}{2}\right)= &
\frac{4}{\pi}\left(V_G - \frac{V_0}{2}\right) \nonumber\\
&\sum_{m=0}^{\infty}
\frac{(-1)^m}{(2m+1)\, I_0(\alpha_{2m+1} a)},
\end{align} 
with $n=2m+1$.
%
Here, $I_0(\alpha_{2m+1} a)\sim e^{\alpha_{2m+1} a}$ results in the higher terms for $m\gg 1$ ignored, and for the lower values of $m$ considering a long pore ($\kappa l \gg1$), we approximate $\alpha_{2m+1}\approx \kappa$. 
Using $\sum_{m=0}^{\infty} \frac{(-1)^m}{(2m+1)}=\pi/4$, Eq.~\eqref{eq:zeta-dirichlet-explicit} at the center of the pore $z=l/2$ simplifies to Eq.~\eqref{eq:zeta_eq_longpore_finiteEDL}. 

Also, using  $\int_0^l \sin\!\left(\frac{n\pi z}{l}\right)\, dz
=
l(1 - (-1)^n)/(n\pi)$ and $\sum_{m=0}^{\infty} \frac{1}{(2m+1)^2}
= \frac{\pi^2}{8}$, we get
\begin{align}
    \sum_{n=1}^{\infty}
\frac{b_n}{I_0(\kappa a)}
\int_0^l \sin\!\left(\frac{n\pi z}{l}\right)\, dz
 = \frac{l(V_G - V_0/2)}{ I_0(\kappa a)}.
\end{align}
Which simplifies the integral $\int_0^l \zeta_{\mathrm{eq}}(z)\, dz$ within $\lambda \rightarrow 0$ limit for long pores ($\kappa l \gg 1$ and hence $\alpha_n \approx \kappa$) from
\begin{equation}
\int_0^l \!\! \zeta_{\mathrm{eq}}(z) dz
=
\!\frac{
\int_0^l \!\left(V_G - \frac{V_0 z}{l}\right) dz
-\!
\sum_{n=1}^{\infty}\!
\frac{b_n}{I_0(\kappa a)}
\!\int_0^l\! \sin\!\left(\frac{n\pi z}{l}\right) dz}{1 - \dfrac{1}{I_0(\kappa a)}}\nonumber
\end{equation}
to $l\left(V_G - \frac{V_0}{2}\right)$ resulting in an average zeta potential expressed in Eq.~\eqref{eq:zeta_av}.

Similarly, the axial electric field from Eq.~\eqref{eq:axial_Ez} can be approximated for a long pore using, 
\begin{align}
  \sum_{n=1}^{\infty} \frac{n\pi}{l} \, b_n \cos\!\left(\frac{n\pi z}{l}\right) = -V_0/l
\end{align} into Eq.~\eqref{eq:Ez_finiteEDL_longPore}.
\bibliographystyle{cas-model2-names}

\bibliography{cas-refs}



\end{document}